\documentclass[sigconf, 10pt]{acmart}
\usepackage{url}
\usepackage{tabularx}
\usepackage{subfigure}
\usepackage{booktabs}
\usepackage{color}
\usepackage{xspace}
\usepackage{comment}

\newcommand{\sn}{\textbf{\textit{N$^2$LoS}}\xspace}

\AtBeginDocument{%
  \providecommand\BibTeX{{%
    \normalfont B\kern-0.5em{\scshape i\kern-0.25em b}\kern-0.8em\TeX}}}

\setcopyright{acmcopyright}
\copyrightyear{2024}
\acmYear{2024}
\acmDOI{XXXXXXX.XXXXXXX}

\acmConference[Conference acronym 'XX]{Make sure to enter the correct
  conference title from your rights confirmation email}{June 03--05,
  2018}{Woodstock, NY}
\acmPrice{15.00}
\acmISBN{978-1-4503-XXXX-X/18/06}

\renewcommand\footnotetextcopyrightpermission[1]{} % removes footnote with conference information in first column
\begin{document}
\newcolumntype{C}[1]{>{\centering\arraybackslash}p{#1}}  
\hyphenation{mm-Wave}
\title{\sn: Single-Tag mmWave Backscatter for Robust Non-Line-of-Sight Localization}

\author{Zhenguo Shi}
\email{zhenguo.shi@unsw.edu.au}
\affiliation{%
  \institution{University of New South Wales}
  \city{Sydney}
  \state{NSW}
  \country{Australia}
}

\author{Yihe Yan}
\email{yihe.yan@student.unsw.edu.au}
\affiliation{%
  \institution{University of New South Wales}
  \city{Sydney}
  \state{NSW}
  \country{Australia}
}

\author{Yanxiang Wang}
\email{yanxiang.wang@unsw.edu.au}
\affiliation{%
  \institution{University of New South Wales}
  \city{Sydney}
  \state{NSW}
  \country{Australia}
}

\author{Wen Hu}
\email{wen.hu@unsw.edu.au}
\orcid{0000-0002-4076-1811}
\affiliation{%
  \institution{University of New South Wales}
  \city{Sydney}
  \state{NSW}
  \country{Australia}
}
\author{Chun Tung Chou}
\email{c.t.chou@unsw.edu.au}
\orcid{0000-0003-4512-7155}
\affiliation{%
  \institution{University of New South Wales}
  \city{Sydney}
  \state{NSW}
  \country{Australia}
}

\renewcommand{\shortauthors}{Zhenguo Shi, et al.}

\begin{abstract}

The accuracy of traditional localization methods significantly degrades when the direct path between the wireless transmitter and the target is blocked or non-penetrable. This paper proposes \sn, a novel approach for precise non-line-of-sight (NLoS) localization using a single mmWave radar and a backscatter tag. \sn leverages multipath reflections from both the tag and surrounding reflectors to accurately estimate the target’s position.  
\sn introduces several key innovations. First, we design \textbf{HFD} (Hybrid Frequency-Hopping and Direct Sequence Spread Spectrum) to detect and differentiate reflectors from the target. Second, we enhance signal-to-noise ratio (SNR) by exploiting the correlation properties of the designed signals, improving detection robustness in complex environments. Third, we propose \textbf{FS-MUSIC} (Frequency-Spatial Multiple Signal Classification), a super-resolution algorithm that extends the traditional MUSIC method by constructing a higher-rank signal matrix, enabling the resolution of additional multipath components.  We evaluate \sn using a 24 GHz mmWave radar with 250 MHz bandwidth in three diverse environments: a laboratory, an office, and an around-the-corner corridor. Experimental results demonstrate that \sn achieves median localization errors of \textbf{10.69 cm (X)} and \textbf{11.98 cm (Y)} at a 5 m range in the laboratory setting, showcasing its effectiveness for real-world NLoS localization.

%$5.95$cm for the laboratory, $12.91$cm for the office and $8.87$cm for the corridor configurations, respectively.

%In particular, to effectively utilize a non-linear adaptive filter (NLAF) for BP measurement, we propose a novel algorithm to generate an effective reference signal for NLAF.
\end{abstract}


%\ccsdesc[500]{Human-centered computing~Ubiquitous and mobile computing systems and tools}

%\keywords{mmWave, NLoS, localization, retro-reflector}


\maketitle

\vspace{-0.2cm}
\section{Introduction}

Indoor localization is a crucial enabler for the Internet of Things (IoT), supporting a wide range of applications, including smart homes, healthcare monitoring, and industrial automation. To achieve high-accuracy localization, extensive research has been conducted, leading to various advanced techniques.  
Among them, optical localization has garnered significant attention due to its ability to provide precise and high-resolution positioning \cite{10.1145/3306346.3322937,10.1145/3665139}. These methods primarily rely on imaging data from cameras to determine location; however, they are highly sensitive to environmental conditions and pose privacy concerns due to the use of cameras~\cite{li2023nlost,somasundaram2023role,velten2012recovering}.  

As an alternative, radio frequency (RF) signals have demonstrated notable potential by leveraging reflections and variations in signal propagation \cite{10.1145/3117811.3117840}. Several RF-based localization techniques utilize radio frequency identification (RFID), Bluetooth, WiFi, or ultra-wideband (UWB) \cite{8550815,10.1145/3411834,10.1145/3577927}. However, these methods often suffer from short-range limitations, reduced accuracy, or require complex receiver setups. To address these challenges, recent research has increasingly focused on millimeter-wave (mmWave) signals, which offer superior range and accuracy due to their high frequency and wide bandwidth \cite{bielsa2018indoor,9780203,8682741}.  

Despite their advantages, existing mmWave localization methods are predominantly designed for line-of-sight (LoS) conditions, requiring an unobstructed path between the transmitter and receiver. However, real-world indoor environments contain numerous obstacles such as walls, furniture, and appliances \cite{8515231}, which render the LoS path non-penetrable due to the short wavelength of mmWave signals \cite{zhao2020m,liu2021seirios}. Consequently, mmWave signals undergo reflection and diffraction, leading to significant performance degradation in traditional LoS-based localization systems.  

%\subsection{Challenges in Non-Penetrable NLoS Localization}

To address these limitations, several mmWave-based approaches have been proposed for non-penetrable NLoS localization. MetaSight \cite{10.1145/3498361.3538947} employs multiple metasurfaces to modulate NLoS signals, while PinIt \cite{10.1145/2534169.2486029} relies on numerous reference tags deployed across the environment. Similarly, SuperSight \cite{10.1145/3643832.3661857} utilizes a triangular tag array with three retro-reflective tags for improved localization accuracy.  
While these solutions demonstrate effectiveness, they present several challenges:
\vspace{-1mm}
\begin{itemize}
    \item \textbf{High Deployment Costs:} Many existing approaches require specialized hardware, such as LiDAR, metasurfaces, or dense reference tag arrays, increasing cost and complexity.
    \item \textbf{Limited Performance in Low SNR Conditions:} The weak nature of NLoS reflections makes localization difficult, particularly in environments with low signal-to-noise ratio (SNR).
    \item \textbf{Multipath Interference:} Complex indoor layouts introduce multipath effects that can significantly degrade localization accuracy.
\end{itemize}
\vspace{-1mm}
Thus, achieving accurate localization in non-penetrable NLoS conditions without requiring additional measurement hardware remains a critical challenge, especially in low-SNR and multipath-prone environments.

%\subsection{Proposed Approach: \sn for NLoS Localization}

To overcome these challenges, we propose \sn, a novel scheme that enables accurate NLoS localization using only a single mmWave radar and a backscatter tag. Unlike previous solutions that require multiple reference points, \sn leverages multipath reflections of mmWave signals to estimate the target's position.  

While both mmWave communication modules and mmWave radar-based tags can facilitate localization, the former requires high power consumption due to continuous signal processing and synchronization. In contrast, backscatter-based radar tags operate with minimal power, utilizing simple reflective principles without active data processing, making them ideal for long-term deployment with minimal maintenance.  
Building on these insights, this paper leverages mmWave radar and backscatter tags to achieve power-efficient and high-precision localization.  

Specifically, \sn introduces a novel modulation scheme, \textbf{HFD} (Hybrid Frequency-Hopping and Direct Sequence Spread Spectrum), to accurately distinguish between reflections from the target and the surrounding environment. The core principle of HFD is to modulate the backscatter signal by periodically switching the tag between ``ON'' and ``OFF'' states, creating a distinctive reflection pattern. In the \textbf{ON} state, the tag actively retro-reflects the incident signal via surrounding reflectors, while in the \textbf{OFF} state, it remains passive, reflecting only based on its material properties.  

By adjusting the frequency of these ON/OFF transitions, frequency shift keying (FSK) can be employed to differentiate the tag's scattering from environmental reflections~\cite{soltanaghaei2021millimetro,bae2022omniscatter}. However, FSK is susceptible to frequency offsets that may impact modulation performance~\cite{923814}. While direct sequence spread spectrum (DSSS) improves resilience against frequency offsets, its effectiveness is limited by code length constraints. To overcome these challenges, HFD \textbf{jointly leverages DSSS and frequency hopping spread spectrum (FHSS)}, enhancing reflector detection and improving localization robustness.  

Additionally, we propose \textbf{FS-MUSIC}, an advanced multipath resolution algorithm that extends traditional MUSIC by analyzing signal characteristics across multiple dimensions, rather than relying on single-dimensional information. This multi-dimensional approach significantly improves localization accuracy in complex environments.  
To summarize, the major contributions of this work are as follows:

%\subsection{Summary of Contributions}
\vspace{-4mm}
\begin{itemize}
    \item We propose \sn, a novel mmWave-based localization system designed for non-penetrable NLoS environments, achieving centimeter-level accuracy using only a single tag and mmWave radar.
    \item We introduce the HFD modulation scheme, which jointly employs FHSS and DSSS to distinguish between reflectors and the target while improving SNR without prior environmental knowledge.
    \item We develop FS-MUSIC, an advanced multipath resolution algorithm that enhances localization performance by extracting valuable information from multipath reflections.
    \item We conduct extensive experiments in multiple real-world indoor environments, demonstrating that \sn achieves median localization errors of \textbf{10.69 cm (X)} and \textbf{11.98 cm (Y)} at a 5 m range in the lab setting.
\end{itemize}
\vspace{-2mm}

%The remainder of this paper is organized as follows: Section II presents the system model and methodology, Section III describes the implementation and experimental setup, Section IV evaluates the system performance, and Section V concludes with future research directions.

\begin{comment}

\end{comment}

\section{Background knowledge}
\label{Background knowledge}
\subsection{FMCW signal model}
\sn is designed for mmWave MIMO radar, which uses chirp signals with frequency-modulated continuous-wave (FMCW) waveform $c(t)$: 
% \vspace{-0.2 cm}
\begin{eqnarray}\label{trans}
c(t)=\alpha_Te^{j(2\pi f_ct+\pi\frac{B}{T}t^2)},
\end{eqnarray}
where $\alpha_T$, $f_c$, $B$ and $T$ are, respectively, the amplitude, the starting frequency, chirp bandwidth and the chirp duration.
The transmitted FMCW signal $c(t)$ propagates in the medium and is reflected back to the radar.  Following the processes of mixing and filtering, the intermediate frequency (IF) signal $y(t)$ is given by: 
% \vspace{-0.2 cm}
\begin{eqnarray}\label{IF}
y(t)=\alpha_Re^{j2\pi (f_c\theta+\frac{B}{T}\theta t)},
\end{eqnarray}
where $\alpha_R$ denotes the amplitude of the received signal. The symbol $\theta=\frac{2D}{c}$ denotes the time delay induced by target reflection, 

where $c$ is the speed of light, and $D$ is the distance between the target and the radar. 
$y(t)$ has a frequency of $\frac{B}{T}\theta$, which can be determined by applying Fast Fourier Transform (FFT) to samples of $y(t)$. The distance between the radar and targets can then be determined by using this frequency.

\subsection{Van Atta Array (VAA)}

{VAA plays a viable role in mmWave radar-based localization, which consists of multiple antennas arranged in a mirror-symmetric configuration. This unique architecture makes the phase for the reflected wave a reversal of the incident wave.} As a result, the signal received by each antenna is transmitted through the incident line and is reflected directly back in the direction it came from. Different from phase-conjugated arrays, which rely on active components, VAAs can be built entirely from passive elements and consume little energy, making them promising for real-world localization applications. Millimetro \cite{soltanaghaei2021millimetro} is a switchable VAA design for LoS ranging and identification, which are utilized by \sn.

\section{Problem and solution overview}\label{nlos}

\subsection{Separating reflections from the target and the environment}
\label{Impact of reflector detection}
\begin{figure}[t]
\centering
\includegraphics[width=7.5cm]{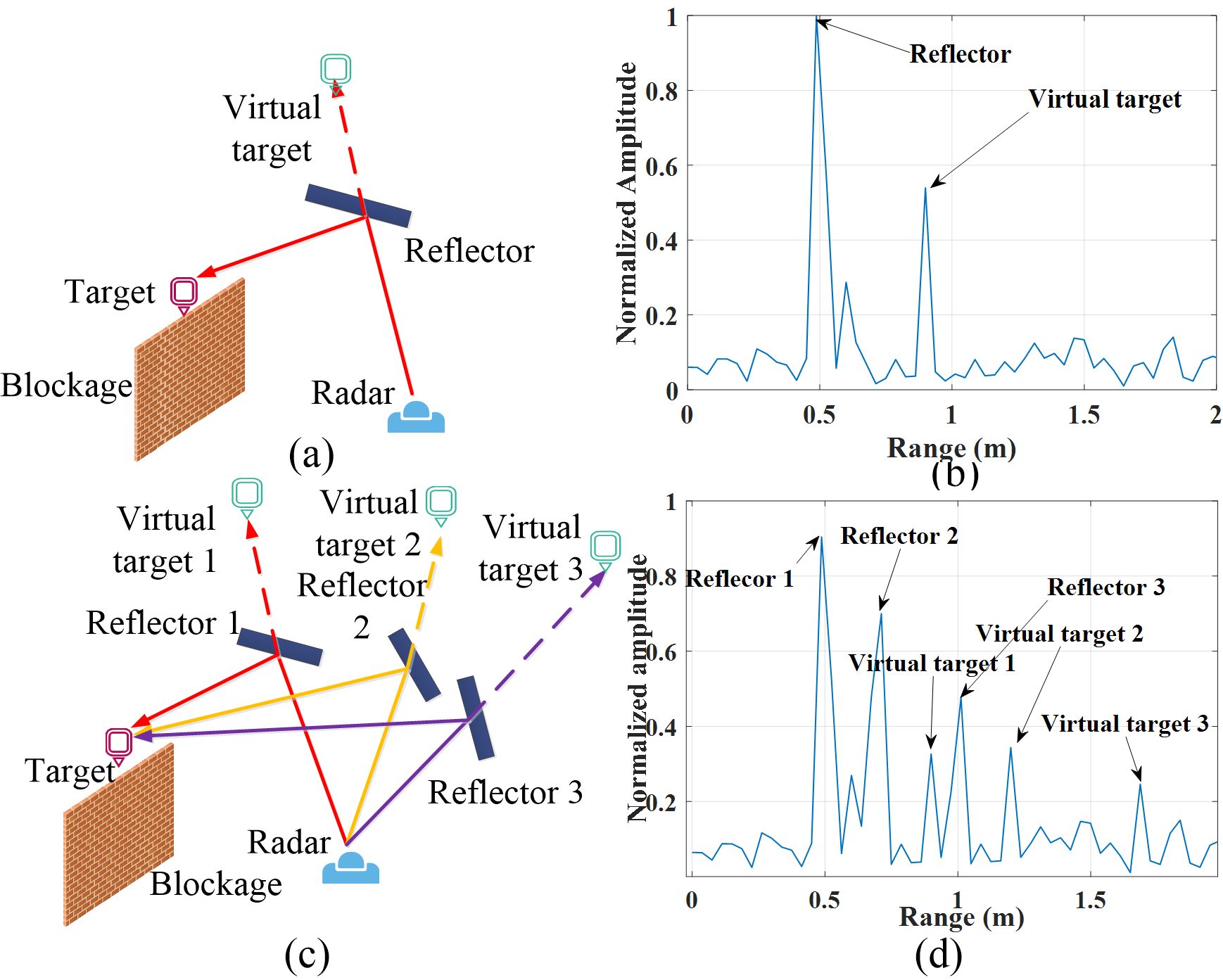}
\DeclareGraphicsExtensions.
\vspace{-0.3 cm}
\caption{Impact of multiple reflectors: (a) and (c) illustrate different reflector layouts, while (b) and (d) show their corresponding range profiles.}
\vspace{-0.8cm}
\label{multi_refelctors}
\end{figure}

For indoor non-penetrable NLoS localization scenarios, multiple reflectors, such as walls and tables, are typically present alongside the target (with a tag). Thus, the radar captures signal peaks from both the reflectors and the target, as shown in Fig.~\ref{multi_refelctors}. The key challenge in such scenarios is to accurately detect and differentiate between reflections from the target and those originating from the surrounding environment.

To address this challenge, significant research efforts have focused on reflector identification and localization. One common approach is to identify the reflector based on the first reflection point, which corresponds to the shortest direct path, while treating the second reflection point as the target (with a single tag), as depicted in Fig.~\ref{multi_refelctors}(a) and Fig.~\ref{multi_refelctors}(b). However, this method is only effective in single-path scenarios and becomes unreliable when multiple reflection paths are present, as shown in Fig.~\ref{multi_refelctors}(c) and Fig.~\ref{multi_refelctors}(d), which is often the case in indoor environments.

%\vspace{-0.2cm}

To overcome this limitation, some studies, such as \cite{10.1145/3517226}, propose using LiDAR for environment profiling and reflector detection. While effective, this approach requires expensive LiDAR devices, leading to high costs and deployment overhead. A recent alternative \cite{10.1145/3498361.3538947} introduces the use of \textbf{multiple retro-reflective tags on the target}, along with multiple radars, to enable reflector and target localization. However, this method imposes strict requirements on tag placement and time synchronization between radars and tags, increasing deployment complexity and costs.

To address these challenges, this work proposes a novel method for reflector identification and target localization using \textbf{only a single tag}, eliminating the need for environmental profiling. This approach significantly reduces deployment costs and complexity while maintaining accurate localization performance in NLoS scenarios.

\subsection{The need to boost the SNR of the returned signal}
\label{Impact of noise on mmWave reflections}
\begin{figure}[t]
\vspace{-0.2 cm}
\centering
\subfigure[Range profile in LoS]
{
\label{snr_los}
\begin{minipage}{0.22\textwidth}
\includegraphics[scale=0.27]{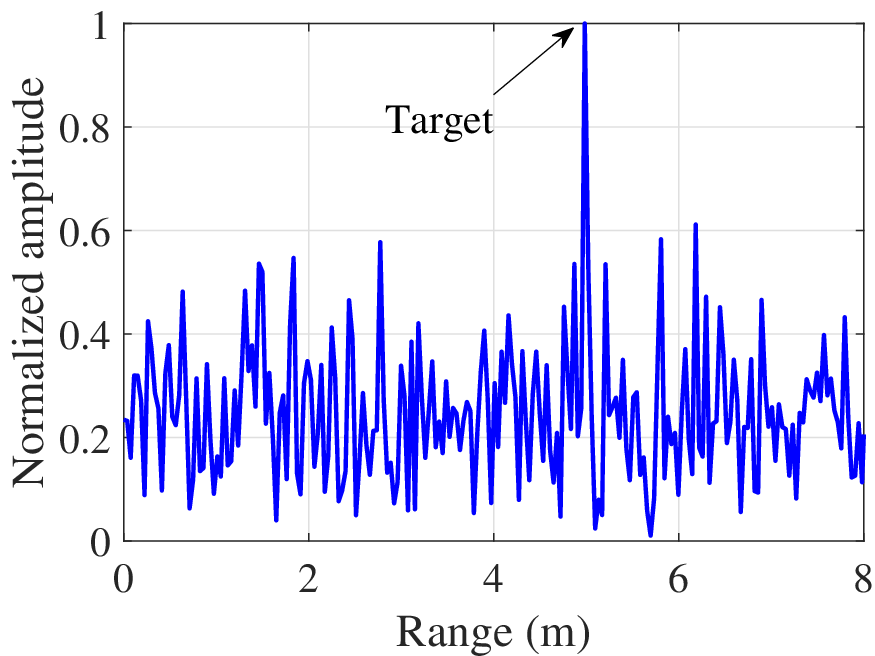}
\end{minipage}
}
\subfigure[Range profile in NLoS ]{
\label{snr_nlos}
\begin{minipage}{0.22\textwidth}
\includegraphics[scale=0.27]{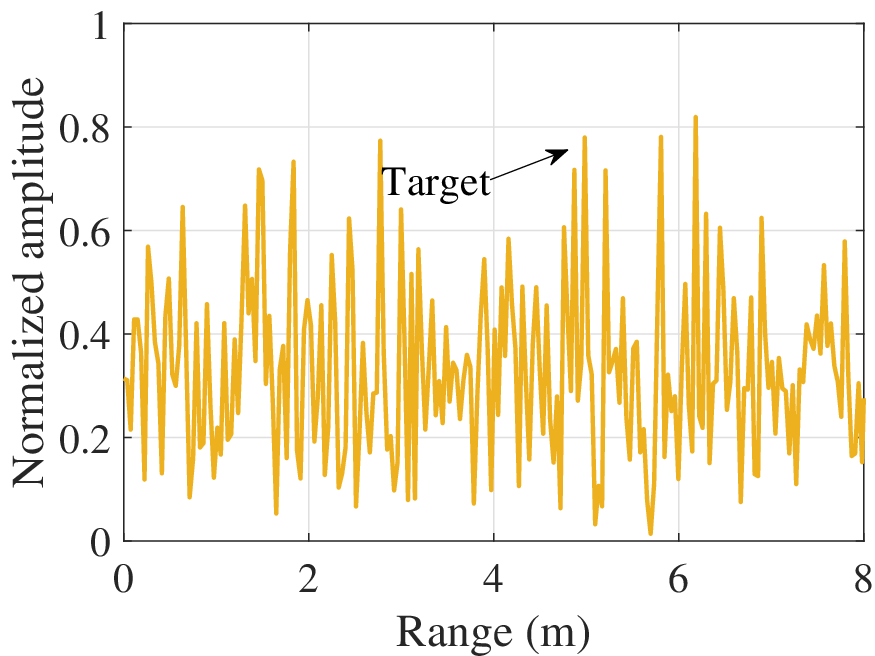}
\end{minipage}
}
\DeclareGraphicsExtensions.
\vspace{-0.3 cm}
\caption{Impact of Low SNR on localization, the distance between radar and target is 5m.}
\vspace{-0.5 cm}
\label{low_snr}
\end{figure}

Indoor non-penetrable NLoS localization often suffers from low SNR due to various environmental factors. Figure~\ref{low_snr} presents the localization results of a single-tag target using FFT in both LoS and NLoS scenarios. In the LoS scenario, the target is effectively localized. However, in the NLoS scenario, the target's position is obscured by noise due to low SNR, preventing accurate localization.

Several factors contribute to the low SNR in NLoS conditions. First, mmWave signals experience significant propagation attenuation as they reflect off walls, furniture, and other obstacles, leading to a considerable reduction in signal strength. Second, reflectors typically absorb a portion of the signal energy during reflection, further reducing SNR. Third, multipath effects introduce interference and noise, causing signal distortion and additional SNR degradation. As a result, low SNR presents a major challenge for accurate target detection and localization in indoor NLoS environments.

To address these challenges, various techniques have been explored to improve SNR through tag modulation, such as FSK and coding schemes. For example, the authors in \cite{8700837} propose an FSK-based approach to enhance SNR and improve target localization accuracy. Similarly, in \cite{482125}, researchers employ DSSS to mitigate interference and boost SNR, thereby enhancing localization performance.
Despite these advances, several limitations still hinder the effectiveness of existing approaches. FSK-based methods are particularly susceptible to frequency-selective fading caused by multipath propagation, resulting in variations in received signal strength and phase, which degrade SNR improvement. Moreover, DSSS-based techniques rely on the length of the code sequence, which directly impacts system performance. While longer codes can accommodate more targets, they also increase system complexity and deployment costs. Given these challenges, this work aims to develop a novel approach that enhances SNR while maintaining scalability for multiple targets.

%\vspace{-0.8cm}
\subsection{The need to improve the number of resolved multi-paths}
In non-penetrable NLoS scenarios, signal reflections from multipath propagation provide valuable localization information. Conventional MUSIC-based methods have shown promise in extracting localization data from multipath signals. However, their performance is strongly influenced by the rank of the covariance matrix of received signals, where a higher rank enables the resolution of a greater number of multipath components. The rank of the covariance matrix is primarily determined by the number of antennas or the spatial degrees of freedom \cite{10.1145/2971648.2971665}.
Due to cost constraints, commercial radars often have a limited number of antenna pairs, such as $2$ transmit (Tx) $\times$ $3$ receive (Rx) or $3$ Tx $\times$ $4$ Rx configurations, which restrict their resolution capabilities. While increasing the number of antennas can improve performance, it also introduces higher financial costs, increased hardware complexity, and energy losses due to signal propagation, all of which negatively impact localization accuracy.

To overcome these limitations, this work proposes increasing the measurement degrees of freedom in the spatial-frequency domain. By leveraging additional frequency diversity alongside spatial information, the effective rank of the covariance matrix can be enhanced, allowing the resolution of more multipath components even with a small number of antenna pairs.

% XXX No figures? Where are the experimental results? 

\vspace{-0.4cm}
\section{Design of \sn }
\label{Design of sn}
%In this section, we first provide the overview of \sn system, then detail the major stages accordingly. 

\subsection{\sn overview}
\begin{figure}[htb]
\centering
\includegraphics[width=\linewidth]{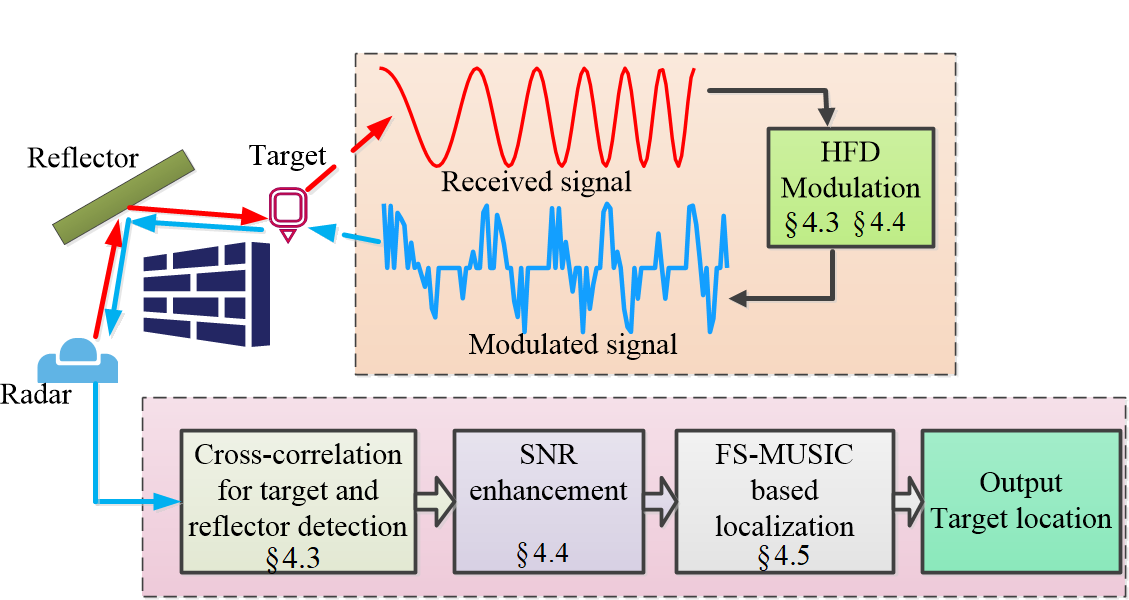}
\DeclareGraphicsExtensions.
\vspace{-0.8 cm}
\caption{System overview}
\label{system}
\vspace{-0.5 cm}
\end{figure}

\sn is designed to provide a precise and practical mmWave-based localization system for non-penetrable NLoS scenarios. Fig.~\ref{system} shows the localization process of \sn.
First, the radar transmits FMCW chirp signals for localization. These signals reach the target with a single tag through reflections from surrounding reflectors. Upon receiving the signals, the tag performs HFD modulation using a novel sequence. 

The radar then captures the modulated HFD signals from the tag (mounted on the target) based on the retro-reflection principle. To accurately distinguish the target from reflectors, \sn utilizes two key components of HFD: the target localization code (TLC) and the reflector localization code (RLC). The TLC is designed to mitigate noise and enhance the SNR of the retro-reflected signal by leveraging DSSS and FHSS techniques. Meanwhile, the RLC identifies reflectors while minimizing localization power consumption.

Next, the FS-MUSIC algorithm is employed to extract multipath reflection features processed by HFD, further improving localization accuracy. Finally, \sn determines the target’s position, providing precise localization results.

%In particular, we exploit the correlation property of TLC to mitigate the impact of noise and enhance SNR levels.
\vspace{-0.3 cm}

\subsection{NLoS Localization with \sn}
\label{Overview:Problem_formulation}
Indoor localization presents a more formidable challenge than its outdoor counterpart due to the cluttered indoor environment, which includes obstacles and reflectors such as walls, doors, and furniture. In typical indoor scenarios, the propagation paths of mmWave radar radio signals between the radar and the localization target are often NLoS and non-penetrable because mmWave radio signals cannot penetrate obstacles. The NLoS propagation means that the transmitted radar signal reflects one or more times before reaching the target, and the same holds true for the return path from the target to the radar. Consequently, a key challenge in indoor localization is extracting useful information from these reflected paths. In this work, we focus exclusively on first-order reflected paths which have higher power levels compared to second or higher-order reflected paths \cite{10.1145/3550320}.
\begin{figure}[t]
\centering
\includegraphics[width=5.5cm]{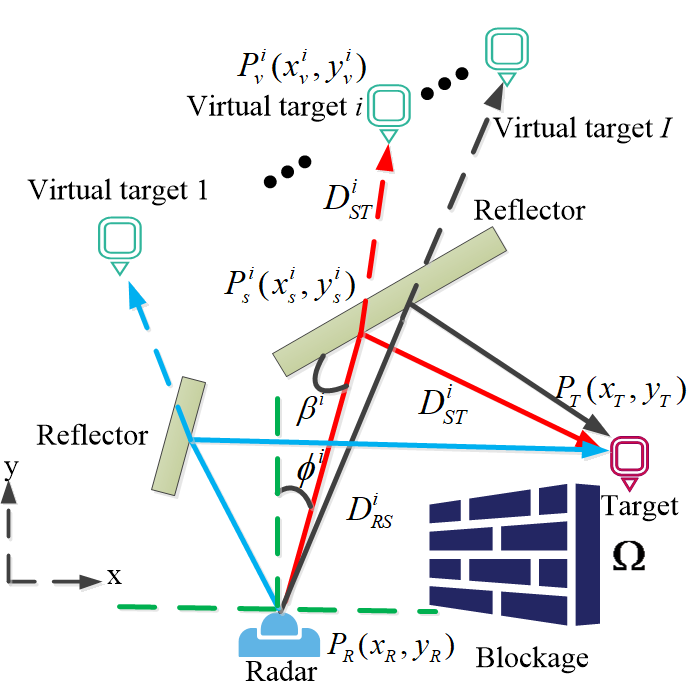}
\DeclareGraphicsExtensions.
\vspace{-0.3 cm}
\caption{A typical NLoS scenario for \sn.}
\vspace{-0.5 cm}
\label{layout_NLOS}
\end{figure}

Fig.~\ref{layout_NLOS} illustrates a typical indoor NLoS localization scenario that is addressed by \sn. The scenario consists of a radar, a target with a single tag, and multiple reflectors which can be walls and furniture. We assume that the direct path between the radar and the target is blocked by an obstacle. We aim to use the reflected signal received by the radar to localize the target. To enhance the signal strength of the radar received signal, we embed a VAA into the tag (on the target) so that it retro-reflects the radar signal. Fig.~\ref{layout_NLOS} shows three NLoS paths (purple, red, and yellow) from the radar to the tag/target via different reflectors. We now present a geometric analysis to explain how \sn uses a number of \textit{key geometric quantities} (specifically distances and angles) to estimate the location of the target. 

Let $P_R(x_R,y_R)$ and $P_T(x_T,y_T)$ denote, respectively, the positions of the radar and the target where $x_R$, $y_R$, etc., are coordinates. We consider $I$ NLoS multipath paths where $I = 3$ in Fig.~\ref{layout_NLOS}. The $i$-th NLoS path, highlighted in red, is shown as a representative example. Specifically, the red path hits the reflector at $P_s^i(x_s^i,y_s^i)$ which refers to the $i$-th reflection point. As depicted in Fig.~\ref{layout_NLOS}, the radar signal goes from $P_R\to P_s^i\to P_T$, and then retro-reflected by the target and returns to the radar along the reverse path $P_T\to P_s^i\to P_R$.

The received signal at the radar consists of two components: the signal reflected solely by the reflectors ($P_R\to P_s^i\to P_R$) and the signal retro-reflected by the target ($P_R\to P_s^i\to P_T\to P_s^i\to P_R$). Later sections of this paper will explain how \sn isolates these $2I$ signal components. Here, we outline the geometric quantities that \sn can extract.  
\sn utilizes the signal along the path $P_R\to P_s^i\to P_R$ to determine $D_{RS}^i$, the distance between the radar at $P_R$ and the reflection point at $P_s^i$, as well as its angle of arrival (AoA) $\phi_i$, as illustrated in Fig.~\ref{layout_NLOS}.  

It is important to note that the reflection signal ($P_R\to P_s^i\to P_R$) results from scattering reflection, which occurs even on smooth and flat surfaces due to microscopic surface roughness or imperfections \cite{balanis2024balanis}. These irregularities cause a portion of the signal to scatter in multiple directions, including back toward the radar \cite{tsang2000scattering}. In real-world environments, surfaces such as walls, furniture, and metallic objects are rarely perfectly smooth, making scattering a common phenomenon. These scattered signals play a crucial role in determining the geometric relationship with reflectors and are widely utilized in mmWave radar localization systems \cite{kong2024survey}. \sn uses this same mechanism to localize reflectors.

Next, we consider the target reflection, where a single tag modulates and reflects the incoming radar signal, generating a strong and distinguishable return. From the radar's perspective, this retro-reflection creates the effect of the target appearing at its mirror image (virtual) position relative to the reflector.

For the red path in Fig.~\ref{layout_NLOS}, the virtual target is located at the point $P_v^i$ located on the continuation of the directed line $P_R\to P_s^i$ such that the distances $|P_s^i P_v^i|$ and $|P_s^i P_T|$ are equal; note that this is a consequence of the Law of Reflection which says that the angles of incidence and reflection are equal. For \sn, it uses the retro-reflection along the red path to obtain $D^i$, which is the distance between the radar and the Virtual target $i$, and its AoA, which is again $\phi_i$. Note that $D^i=D^i_{RS}+D^i_{ST}$ where $D^i_{ST}$ is the distance between reflection point $P^i_s$ and the target. 

We now explain how \sn uses the key geometric quantities $\{ D^i, D_{RS}^i, \phi_i \}_{i = 1, \ldots, I}$ to obtain the position of the target. First, \sn computes the coordinates of reflection point $P_s^i$:
\begin{eqnarray}\label{ps}
x_s^i=x_R+D^i_{RS}\cos{(\phi^i)},
\quad{}
y_s^i=y_R+D^i_{RS}\sin{(\phi^i)}.
\end{eqnarray}

Next, we use the fact that the distance from the target to the reflection point $P_s^i$ is $D^i_{ST}$ which can be calculated from $D^i-D^i_{RS}$. We can therefore solve a multilateration problem to obtain the position of the target using the calculated positions of $\{ P_s^i \}_{i = 1,\ldots, I}$ as the anchors and $D^i_{ST}$ as the distances of the target to the anchors \cite{pandey2006survey}. 

The above description shows that \sn does \textit{not} require any knowledge on the geometry of the environment, which is an advantage. It also shows that \sn needs 3  assumptions to work. First, it assumes that every NLoS path passes through only one reflector. This is justifiable as higher-order reflected paths have lower energy \cite{10.1145/3550320} and are not detectable. Next, two assumptions are needed so that the multilateration problem returns a unique tag position: there are at least 3 reflection points to act as anchors, and these points are not collinear. Since each NLoS path gives a reflection point and the indoor environment is known to be rich in NLoS paths \cite{10.1145/3498361.3538947}, so there are generally many reflection points. Also, these reflection points come from walls and furniture, so they are unlikely to be collinear. 

Since the key geometric quantities required by \sn are the distances from the radar to the reflection points and the virtual targets, as well as the AoA's, this suggests that \sn needs to: (1) separate the reflections from the reflector and the target; (2) improve the signal power of the retro-reflection as it takes a longer path; and (3) ensure that sufficient multi-paths can be resolved for the multilateration problem. We will dwell deeper into these requirements next. 
\vspace{-0.5 cm}
\subsection{HFD based target and reflectors detection}\label{RD}
As discussed in Section \ref{nlos} above, NLoS localization requires accurate information of reflector positions. While theoretical models often assume reflector positions are known, real-world deployment faces challenges due to the dynamic and unpredictable nature of indoor environments, where reflector locations may frequently change. Prior works attempt to obtain the reflector information by employing LiDAR \cite{10.1145/3517226} or dense reference tags \cite{10.1145/3643832.3661857}. However, these methods generally increase system complexity and operational costs. To address these challenges, we propose a novel modulation method, termed HFD, to accurately detect reflector positions, providing essential information for successful localization.

\begin{figure}[htb]
\centering
\includegraphics[width=8.2 cm]{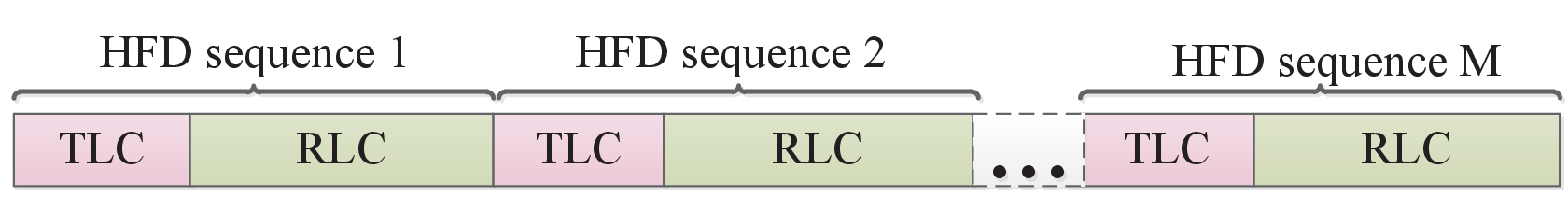}
\DeclareGraphicsExtensions.
 \vspace{-0.4 cm}
\caption{The Structure of HFD Sequence}
\label{new_HFD_structure}
 \vspace{-0.6 cm}
\end{figure}

The primary objective of HFD is to modulate the signals received at the tag (on the target), enabling the distinction between the \textit{modulated} signals from the tag/target and \textit{unmodulated} signals reflected by walls,  surroundings, etc. Fig.~\ref{new_HFD_structure} shows that  HFD consists of cycles of target localization code (TLC) and reflector localization code (RLC) phases. The high-level difference between these two phases is that during TLC, the tag on the target modulates received radar signals before retro-reflecting it; however, during RLC, the tag does \textit{not} modulate or retro-reflect. 

%The RLC is designed to identify reflectors and their surroundings, enabling their differentiation from the target while minimizing target power consumption. 

In particular, in the TLC phase, the tag is active and modulates signals following an On/Off switching operational principle, resulting in modulated signals with high auto-correlation and low cross-correlation. Subsequently, at the radar receiver, the incoming signal is cross-correlated with the corresponding designed modulation signals, leveraging their auto-correlation properties to achieve target detection. In contrast, in the RLC phase, the tag remains inactive and reflects signals solely based on its material properties. It should be noted that the reflected signal from the tag in the RLC state is always too weak to detect, primarily due to the NLoS conditions typical in indoor environments, where the signal is overwhelmed by noise. As a result, the radar only detects signals reflected directly by the reflectors and surroundings but not any signal from the tag. Note that, the tag in RLC state remains in an inactive "Off" state, minimizing power consumption by eliminating active signal transmission and processing. By utilizing both TLC and RLC, we can effectively differentiate the tag (equipped on the target) from the reflectors and surroundings. 

\begin{figure}[htb]
\centering
\includegraphics[width=\linewidth]{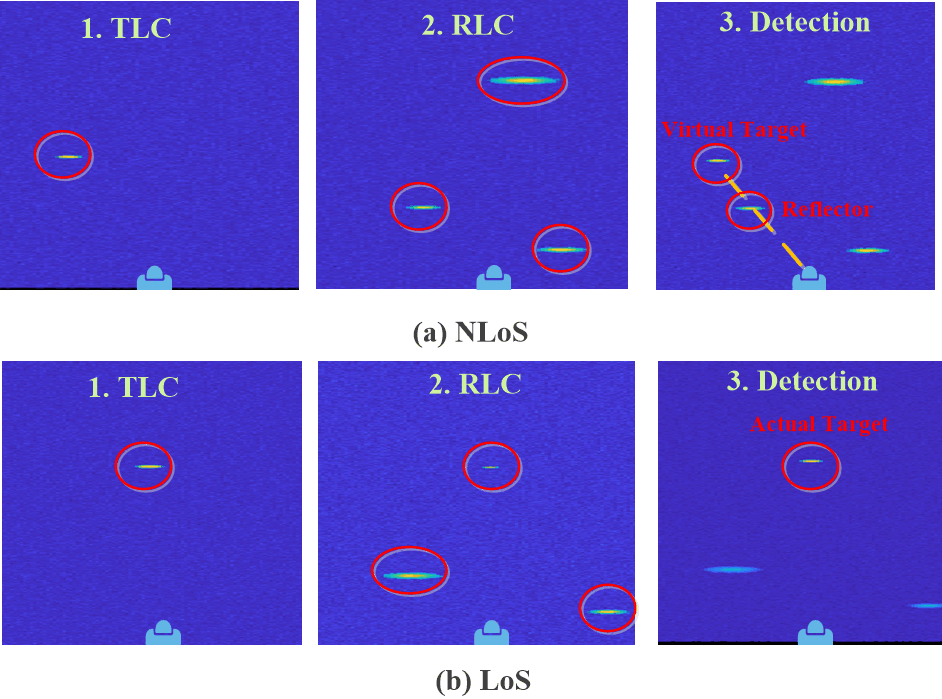}
\DeclareGraphicsExtensions.
\vspace{-0.8 cm}
\caption{Target and Reflector Detection.}
\vspace{-0.5 cm}
\label{detection}
\end{figure}

We now use Fig.~\ref{detection} to explain the inference that we can make from the radar signals received during the TLC and RLC phases. Fig.~\ref{detection}(a) shows an NLoS scenario where a virtual target and reflectors are located, respectively, in TLC and RLC; while Fig.~\ref{detection}(b) shows the result when the target is in the LoS of the radar. Note that the range and AoA of the virtual target are obtained by the FS-MUSIC in Section~\ref{MCS}, while the reflector locations can be obtained via the MUSIC algorithm and \eqref{ps}. We want to illustrate two lessons here. First, we can use these results to tell whether a scenario has a LoS path or not. If the target is in the LoS of the radar, as in Fig.~\ref{detection}(b), the target will appear in both TLC and RLC at approximately the same location. However, in the non-penetrable NLoS scenario in Fig.~\ref{detection}(a), the target only appears in TLC but not RLC. {
By following these steps, the LoS or NLoS condition can be determined by comparing the target's presence in the TLC and RLC phases. The effectiveness of HFD in distinguishing between LoS and NLoS conditions will be evaluated in Section~\ref{relector_detection}.  
}
Second, we can see in the right-most pane in Fig.~\ref{detection}(a) that, in the NLoS scenario, the virtual target detected in TLC is on the directed line extending from the radar to one of the reflectors. This confirms the depiction in Fig.~\ref{multi_refelctors}. Note that there is an angular difference between the lines connecting the virtual target and reflector. In this paper, an angular difference within 1 degree is considered to align the virtual target along the same line with the reflector, treating them as collinear points for localization purposes.

\subsection{HFD based SNR improvement} \label{HFD_snr}
As discussed earlier, HFD is designed to achieve accurate localization. However, in indoor environments, mmWave signals undergo path attenuation, scattering, absorption, and reflection due to obstacles and reflectors before reaching the radar receiver. Consequently, the received signal energy at the radar, $r_{rad}$, is significantly reduced, resulting in an SNR that is too low for precise localization. While conventional methods, such as filtering, can partially enhance SNR, additional techniques are required to further improve performance.

To address these challenges, we propose a TLC to modulate the received signal at the tag before retro-reflecting it, leveraging both DSSS and FHSS. Specifically, TLC employs FHSS to hop across different frequency channels, while DSSS modulation is applied within each frequency hop. DSSS operates by switching the signal state, where an ``on'' state represents a binary 1 and an ``off'' state represents a binary 0. FHSS extends this by dynamically adjusting the switching frequency, enabling adaptive frequency hopping. This combined approach enhances localization accuracy while overcoming DSSS's user support limitations and reducing power consumption, which will be evaluated in Section~\ref{Power Consumption}.

Let $r_e(t)$ denote the radar signal received at the tag (on the target), and $g(t)\in \{1, 0\}$ stands for a DSSS sequence at time $t$. Then, the DSSS modulated signal $r_d(t)$ is obtained by
\begin{eqnarray}\label{r_d}
r_d(t)=r_e(t) \cdot g(t),
\end{eqnarray}

Next, FHSS is employed to vary the carrier frequencies of DSSS sequences. Let $ \{f_1, f_2, \ldots, f_{N_f}\} $ be the set of available frequency channels. The frequency hopping sequence $f_H(t)$ is determined by a pseudorandom index function $ind(t)$, which maps each time $t$ to one of the frequency channels. Consequently, the carrier frequency at time $t$ is obtained by $f_H(t) = f_{ind(t)}$, where $ind(t)$ is a pseudorandom function generating indices within the set ${1, 2, \ldots, N_f}$. Then, the tag signals are modulated as 
\begin{eqnarray}\label{r_m}
r_{m}(t) = [r_e(t) \cdot g(t)] \cdot \cos(2\pi f_H(t) t)
=r_e(t)\cdot g_{T}(t),
\end{eqnarray}
where $g_{T}(t)= g(t) \cdot \cos(2\pi f_H(t) t)$ stands for the TLC sequence. Next, the modulated signal $r_{m}(t)$ is retro-reflected to radar via the reflectors. 

At the radar side, after applying mixing, filtering and dechirping, the received base-band signal $r_{rad}(t)$ can be expressed as:
\begin{eqnarray}\label{r_rad}
r_{rad}(t) =r_{target}(t) + r_s(t)+w(t),
\end{eqnarray}
where $r_s(t)$ stands for the reflected signals from reflectors and other surroundings, and $w(t)$ is the noise. $r_{target}(t)$, the retro-reflected signals from the target, is a combination of $I$ reflected signals passing through multiple paths:
\begin{eqnarray}\label{r_tag}
r_{target}(t) =\sum_{i=1}^I \alpha^i g(t)e^{j2\pi (f_H(t)+\Delta^i)t},
\end{eqnarray}
where $\alpha^i$ represents the attenuation of $i$th path, and $\Delta^i$ is the frequency shift due to the distance between the radar and target in the $i$th path. In typical indoor environments, the energies of $r_s$ and $w$ are relatively high, whereas $r_{target}$ exhibits lower energy levels due to longer path attenuation and signal energy absorption by reflectors. This results in a low SNR condition, posing challenges for precise localization.

To address the above concern, we propose to leverage the correlation property of TLC to eliminate the noise and interference, whilst retaining the modulated target/tag signals. As demonstrated in (\ref{r_rad}) and (\ref{r_tag}), only $r_{target}$ incorporates TLC modulation, resulting in a strong correlation with $g_T$, whereas $r_s$ and $w$ exhibit weak correlations with $g_T$. By utilizing this property, we can effectively enhance $r_{target}$, while minimizing the interference from $r_s$ and $w$. To this end, a sliding correlation will be conducted on $g_T$ and received signal, expressed as
\begin{eqnarray}\label{sli_cor}
r(k)=\!\!\sum_{n}^{N_T}\!r_{tag}(n)g_T(n\!+\!k)+ \sum_{n}^{N_T} (r_{s}(n) + w(n))g_T(n\!+\!k).\!\!\
\end{eqnarray}
%\nonumber
%&&+\!\!\sum_{n}^{N_T}w(n)g_T(n\!+\!k\!).
Due to the minimal correlation between $r_s$, $w$, and $g_T$, and the high autocorrelation of $g_T$, the second sum in (\ref{sli_cor}) is approximately zero while the first sum is significantly enhanced, considerably improving the SNR level of $r$. Moreover, based on (\ref{r_tag}) and (\ref{sli_cor}), $r$ retains the target position information ($\Delta_i$), enabling effective localization.

{
It is important to note that the above discussion primarily focuses on the case of a single target. However, the proposed TLC method can also be applied to localizing multiple targets. In a multi-target scenario, each target is equipped with a unique tag that transmits a distinct TLC sequence. These TLC sequences consist of different DSSS and FHSS codes, carefully designed to exhibit strong auto-correlation properties while maintaining minimal cross-correlation with other sequences.  
By performing cross-correlation between the received signal and the corresponding TLC sequence, the reflected signal from each target can be uniquely identified and separated from interference caused by other targets and surrounding objects. This enables the simultaneous detection and localization of multiple targets.

}

\subsection{FS-MUSIC design for target localization}\label{MCS}
\subsubsection{Virtual target localization}
Upon obtaining the enhanced signal reflections $r$ from the previous section, the critical task is to process it to extract valuable features for localization. To achieve this, FMCW signal characteristics are commonly used to determine a target's position by estimating distance and angle through the range and angular FFT analysis  \cite{michel1992localization}. However, this approach is constrained by the radar system's bandwidth and the number of antennas. To mitigate these limitations, super-resolution techniques, e.g., ESPRIT and MUSIC, have emerged as effective solutions \cite{li2016dynamic}. Although promising, the capability of ESPRIT heavily relies on precise estimation of subspaces and eigenvalues, which are notably susceptible to noise \cite{shahbazpanahi2001distributed}. Moreover, it often experiences significant performance degradation under low SNR conditions. In contrast, MUSIC performs the singular value decomposition (SVD) on the covariance matrix ${\bf{R}}$ to obtain the corresponding eigenvalues and eigenvectors. The number of eigenvalues indicates the quantity of resolvable multipath. 
Moreover, MUSIC effectively separates signal and noise components, making it robust in low SNR conditions. This makes it promising for indoor NLoS localization, especially in environments with significant multipath effects and low SNR.
% Moreover, MUSIC can effectively separate signal and noise components, making it robust against noise and performing well in low SNR conditions. Consequently, it exhibits promise for indoor NLoS localization, particularly in environments characterized by significant multipath effects and low SNR conditions. 

However, the efficacy of traditional MUSIC is highly dependent on the number of measurement dimensions, which correspond to the number of antennas in the radar system. Limited antenna availability due to hardware constraints in low-cost radar systems compromises the effectiveness of the MUSIC algorithm. While some studies have explored leveraging frequency dimension information to enhance MUSIC performance \cite{kotaru2015spotfi}, they are tailored to systems with channel state information (CSI) as input, their effectiveness without CSI input remains unverified. Moreover, they do not consider the potential benefits of modulation-related features. To address this, we propose FS-MUSIC, which effectively leverages signal characteristics provided by the HFD modulation (as discussed in Section \ref{HFD_snr}), to improve MUSIC performance. 

%aiming at joint range-azimuth estimation by utilizing information extracted from both frequency and spatial measurement dimensions, rather than solely spatial dimension in the conventional MUSIC method. 

To facilitate FS-MUSIC, we select $r$ as the input signal, as defined in  (\ref{sli_cor}). For a radar with multiple Tx and Rx antenna pairs, the signal matrix $\bf{r}$ can be formulated as
 	\begin{equation}
		\bf{r} = \begin{bmatrix}
			r_{1,1} & r_{1,2} & \cdots & r_{1,n} & \cdots  & r_{1,N} \\
			% r_{2,1} & r_{2,2} &\cdots &  r_{2,n} & \cdots & r_{2,N}\\
			\vdots  & \vdots  &  \vdots &   \vdots   &    \vdots  &   \vdots \\
			r_{N_a,1} & r_{N_a,2} & \vdots & r_{N_a,n}& \vdots & r_{N_a,N} \\
		\end{bmatrix},
		\label{r}
	\end{equation}
 % 	\begin{equation}
	% 	\bf{r} = \begin{bmatrix}
	% 		r_{1,1} & r_{1,2} & \cdots & r_{1,n} & \cdots  & r_{1,N} \\
	% 		r_{2,1} & r_{2,2} &\cdots &  r_{2,n} & \cdots & r_{2,N}\\
	% 		\vdots  & \vdots  &  \vdots &   \vdots   &    \vdots  &   \vdots \\
	% 		r_{N_a,1} & r_{N_a,2} & \vdots & r_{N_a,n}& \vdots & r_{N_a,N} \\
	% 	\end{bmatrix},
	% 	\label{r}
	% \end{equation}
where $N_a$ and $N$ are the number of antennas and the length of the $r$ respectively. 

Based on the above discussion, MUSIC's ability to resolve multipath signals depends on measurement dimensions. Given this, we propose FS-MUSIC to enhance multipath resolution by creating additional measurement dimensions. To achieve that, we explore FHSS modulation characteristics within HFD and construct new measurement dimensions across hopping frequencies, integrating spatial dimensions from multiple antennas. This creates frequency-spatial domain measurement dimensions, enhancing the multipath resolution of the MUSIC. Consider the $j$th antenna's signal ${\bf{r}}_{j,:}$, and we divide it into $N_f $ segments, with each segment length equal to the hopping period $N_h$. Since ${\bf{r}}_{j,:}$ has already undergone time offset compensation, each segment's data comprises the characteristics of a single FH frequency only. For the $n_f$th segment, the frequency components primarily consist of $f_{n_f} + \Delta$, where $f_{n_f}$ is the frequency of the $n_f$th FH, and $\Delta$ represents the frequency shift due to the distance between the radar and the target. As a result, the data in each segment is independent and contains the target location information. Based on this, with each segment data as the smallest unit, the FS-MUSIC signal matrix 
${\bf{\hat{r}}}$ is: 
\begin{equation}
\bf{\hat{r}}\! =\! \begin{bmatrix}
			r_{1,1} & \!\!\cdots \! \! & r_{1,n} & \!\!\cdots \!\! & r_{1,N_{h}} \\
            r_{1,N_{h}+1}   & \!\!\cdots \!  \!&r_{1,N_{h}+n} & \!\!\cdots \!\! & r_{1,2N_{h}} \\
            \vdots   &  \!\!\vdots  \!\! &   \vdots   &    \!\!\vdots \!\! &   \vdots \\
            r_{1,(\!N_f \!- \!1\!)N_{h} \!+ \!1} & \!\!\cdots  \!\! &r_{1,(\!N_f \!- \!1\!)N_{h} \!+ \!n} & \!\!\cdots \! \!\! & r_{1,N_f N_h} \\
            r_{2,1} & \!\!\cdots \! \! & r_{2,n} & \!\!\cdots\!\!  & r_{2,N_{h}} \\
            r_{2,N_{h}+1} & \!\!\cdots \! \! &r_{2,N_{h}+n} & \!\!\cdots\!\!  & r_{2,2N_{h}} \\
            \vdots   &  \!\!\vdots \! \! &   \vdots   &   \!\! \vdots \!\! &   \vdots \\
			r_{N_a,(\!N_f \!- \!1\!)N_{h} \!+ \!1}  & \!\!\cdots  \! \!&r_{N_a,(\!N_f \!- \!1\!)N_{h} \!+ \!n} & \!\!\cdots  \!\! & r_{N_a,N_f N_h} \\
		\end{bmatrix}, 
		\label{eq:esprit:Xk}
\end{equation}

where $\bf{\hat{r}}\in {\mathbb{C}}^{(N_a N_f)\times N_h}$. Note that the data in any two rows of $\bf{\hat{r}}$ are independent of each other, thus the rank of its covariance matrix $\bf{\hat{R}} =E[\bf{\hat{r}}\bf{\hat{r}}^*]$ increases to $N_aN_f$. Accordingly, the maximum number of resolvable multipaths is $\frac{2N_aN_f}{3}$, which is greater than $\frac{2N_a}{3}$ for conventional MUSIC methods \cite{li2016dynamic}. 

{Furthermore, FS-MUSIC enhances angular resolution by leveraging frequency diversity. Conventional MUSIC relies on the number of antennas $N_a$ for resolution $\frac{\lambda}{2 N_a d}$, where $d$ is the antenna spacing, and FS-MUSIC incorporates $N_f$ independent frequency hops, effectively improving resolution to $\frac{\lambda}{2 N_a d} \cdot \frac{1}{N_f}$. This enables finer angular resolution even with a small antenna array, making FS-MUSIC effective in resource-constrained scenarios. The performance of FS-MUSIC on angular estimation will be discussed in Section~\ref{Effectiveness of MUSIC conjugate}.
%is shown in Fig. \ref{algorithm_cdf_distance_office_MUSIC}.
}

Next, by applying SVD on $\bf{\hat{R}}$, the eigenvalues and eigenvectors are obtained by 
$[{\bf{\hat{U}}}, {\bf{\hat{\Lambda}}}]=\text{SVD}(\bf{\hat{R}}),$
% \begin{eqnarray}\label{SVD}
% [{\bf{\hat{U}}}, {\bf{\hat{\Lambda}}}]=\text{SVD}(\bf{\hat{R}}),
% \end{eqnarray}
where ${\bf{\hat{U}}}$ and ${\bf{\hat{\Lambda}}}$ stand for eigenvector matrix and the diagonal matrix of eigenvalues, respectively. Then, the smallest $I$ eigenvectors are selected to form the matrix  $\mathbf{U}_n$, which is the noise subspace. Finally, the 2D MUSIC spectrum \( P(d, \eta) \) is computed as:
\begin{eqnarray}\label{p}
P(d, \eta) = \frac{1}{\mathbf{a}^H(d, \eta) \mathbf{U}_n \mathbf{U}_n^H \mathbf{a}(d, \eta)},
\end{eqnarray}
where $ \mathbf{a}(d, \eta) $ is the 2D steering vector representing both the distance and direction information of the target. Consequently, FS-MUSIC uses signal subspace eigenvectors to enhance responses at specific distances and directions, reducing noise and interference for accurate target distance and direction estimation.
% Therefore, FS-MUSIC utilizes the signal subspace eigenvectors to enhance signal responses at specific distances and directions, effectively mitigating noise and interference for accurate joint distance and direction estimation.
\subsubsection{From Virtual to Actual Target Localization}
\label{From Virtual to Actual Target Localization}
Following the above steps, we can determine the potential location of the target. In NLoS scenarios, this position corresponds to the virtual target location $P_v$. 
% To find the actual position, Equations (\ref{pt}) to (\ref{ps}) can be utilized for solutions. 
% However, some existing methods \cite{XXX} typically assume some known environment parameter, e.g. the angle $\beta^i$ in Fig.~\ref{}, but this is challenging to achieve in practical applications. Although LS methods can relax the constraint of $\beta^i$, they always assign equal weight to all measurements, leading to inaccurate localization if some measurements are less reliable. 
To find the actual target position, we propose to employ the weighted least squares method (WLS) to solve the multilateration problem \cite{pandey2006survey} 
by appropriately weighting measurements based on their reliability to achieve accurate localization \cite{lin2013new}. To be specific, each measurement is weighted according to signal strength, where higher weights are assigned to measurements with stronger signals and more accurate paths. This weighting strategy allows WLS to reduce the impact of unreliable measurements, improving location estimates without additional computational complexity. Once weights are assigned, the target’s position can be accurately computed by minimizing the total weighted error across all measurements. 

For the pair of $P_T$ and $P_s^i$, the weighted error function of WLS, $E^i(x_T,y_T)$, is defined as
\begin{eqnarray}\label{err_s}
E^i(x_T,y_T)=\omega^i\Big[\sqrt{(x_T-x_s^i)^2+(y_T-y_s^i)^2}-D_{ST}^i\Big]^2,
\end{eqnarray}
where $\omega^i$ stands for the weight factor associated with the $i$th scatter point. The total weighted error function $E_{\text{total}}(x, y)$ can be obtained by summing the weighted errors over all the scatter points, expressed as
\begin{eqnarray}\label{err_all}
E_{total}(x_T,y_T)=\sum_{i=1}^{I}E^i(x_T,y_T).
\end{eqnarray}
Next, the actual target position $(x_T, y_T)$ can be obtained by minimizing $E_{\text{total}}(x_T, y_T)$ \cite{lin2013new}.

{
It is noteworthy that the core principle of WLS, as expressed in (\ref{err_s}), lies in determining an optimal $w_i$ based on the measurement system and the associated error characteristics. Traditional optimization algorithms are often limited by their computational complexity. While deep learning-based methods show potential, they require extensive training data and exhibit sensitivity to environmental variations. To address these limitations, we propose a computationally efficient method for weight assignment that avoids additional overhead. Our approach utilizes the eigenvalues $u_i$ derived from FS-MUSIC, where the magnitude of each eigenvalue corresponds to the signal strength of its respective path. Larger eigenvalues indicate stronger signals and more accurate path estimations. Leveraging this property, we assign weights $w_i$ proportionally to the eigenvalues, prioritizing paths with higher signal reliability, which is
}
\begin{eqnarray}\label{wi}
w_i=\frac{u_i}{\sum_{j=1}^I u_j},
\end{eqnarray}
where $I$ is the number of multipaths. Once $w_i$ is determined, we can obtain the target's position by minimizing (\ref{err_all}).

\section{Performance Evaluation}
\label{Performance Evaluation}
% In this section, we evaluate and analyze the performance of the proposed \sn under various configurations and settings.
\begin{figure}[t]
\centering
\includegraphics[width=8.5 cm]{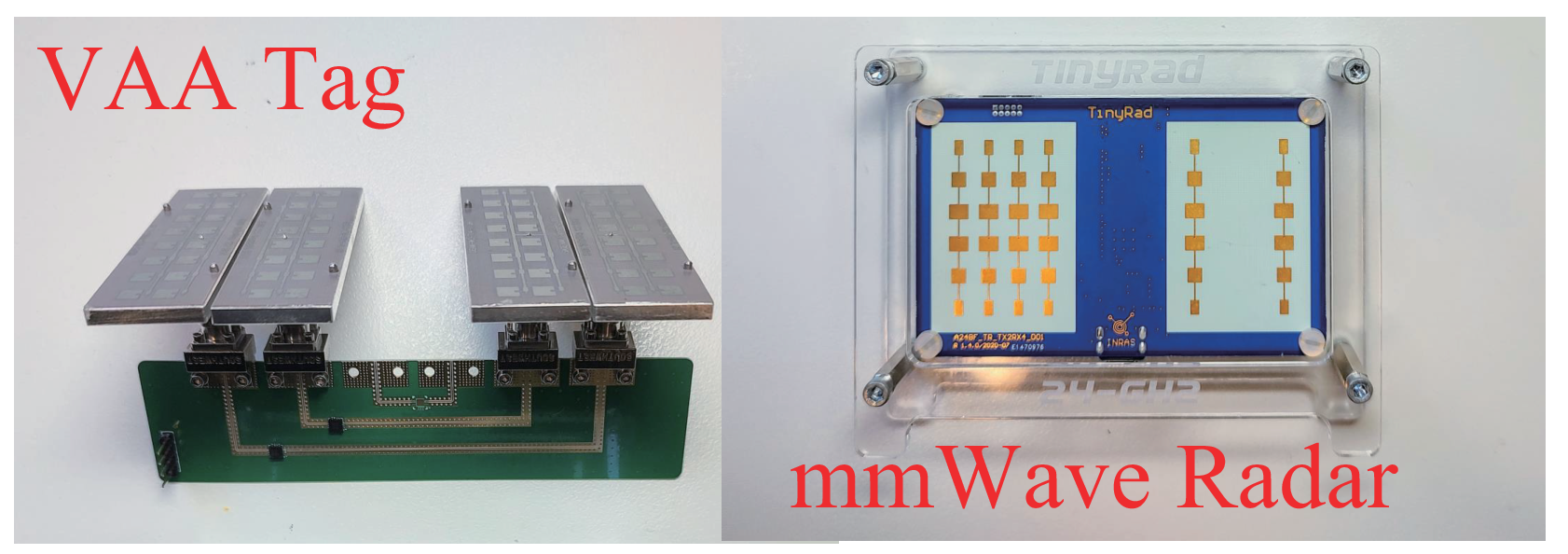}
\DeclareGraphicsExtensions.
\caption{Millimetro~\cite{soltanaghaei2021millimetro} VAA tag (left) and DEMORAD radar (right). }
%\vspace{-0.5 cm}
\label{vaa_radar}
\end{figure}
\begin{figure}[t]
\centering
\includegraphics[width=8.5 cm]{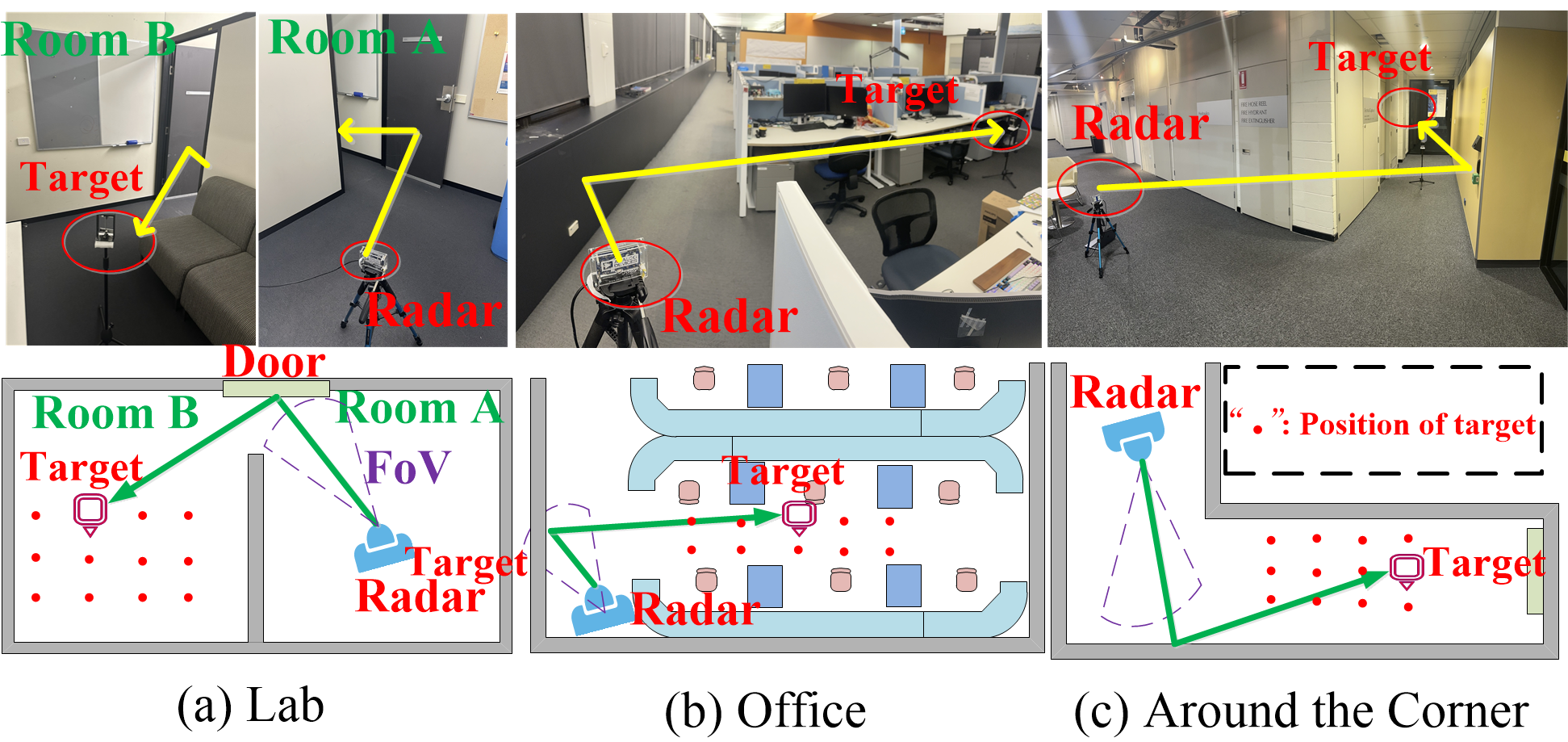}
\DeclareGraphicsExtensions.
\vspace{-0.5 cm}
\caption{Layout of different environment}
% \vspace{-0.5 cm}
\label{experimental_setup}
\end{figure}

\subsection{Experimental Set-up and Metrics}
\label{Experimental Set-up and Metrics}
To assess the the performance of \sn, we develop a prototype (see Fig.~\ref{vaa_radar}) consisting of a COTS 24 GHz DEMORAD radar with a bandwidth of 250 MHz, and a Millimetro VAA tag~\cite{soltanaghaei2021millimetro}, which is paired with an MSP430 microcontroller \cite{mcu2020} and a MACOM RF switch \cite{RFswtich}. The radar platform comprises two transmit and four receive antennas. An enhanced 15-bit Gold code sequence is deployed as DSSS to improve the received signal's SNR and to ensure auto-correlation and cross-correlation properties. For FHSS, the code sequence is modulated at 2 KHz, 5 KHz, and 10 KHz, respectively, and the radar's chirp duration is configured at 2.56 $ms$ with 32 chirps per frame.

\begin{figure*}[t]
\centering
\begin{minipage}{5.7cm}
\centering
\includegraphics[width=1\textwidth]{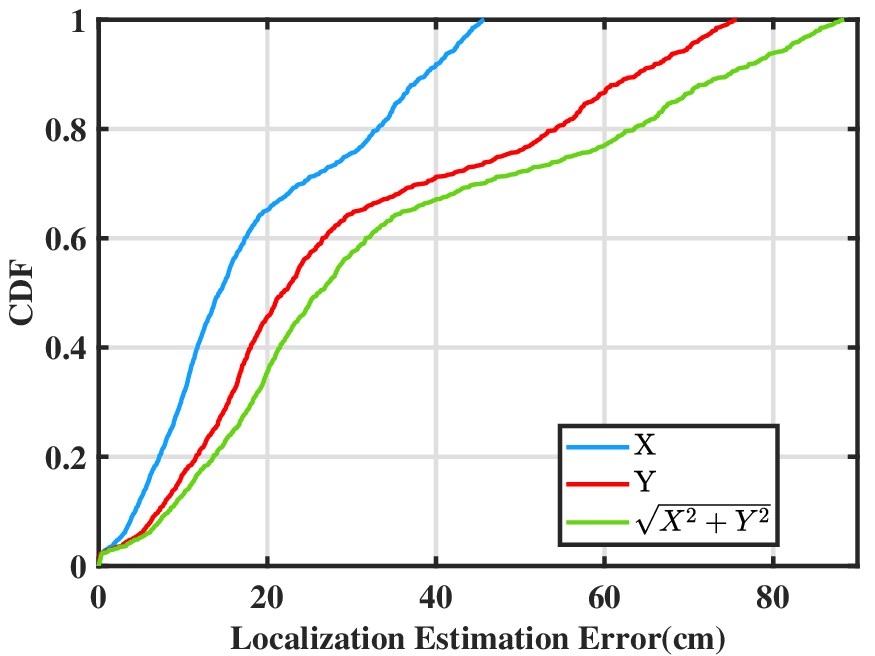}
\caption{Localization error for\\ office.} \label{cdf_distance_office}
\end{minipage}
\begin{minipage}{5.7cm}
\centering
\includegraphics[width=1\textwidth]{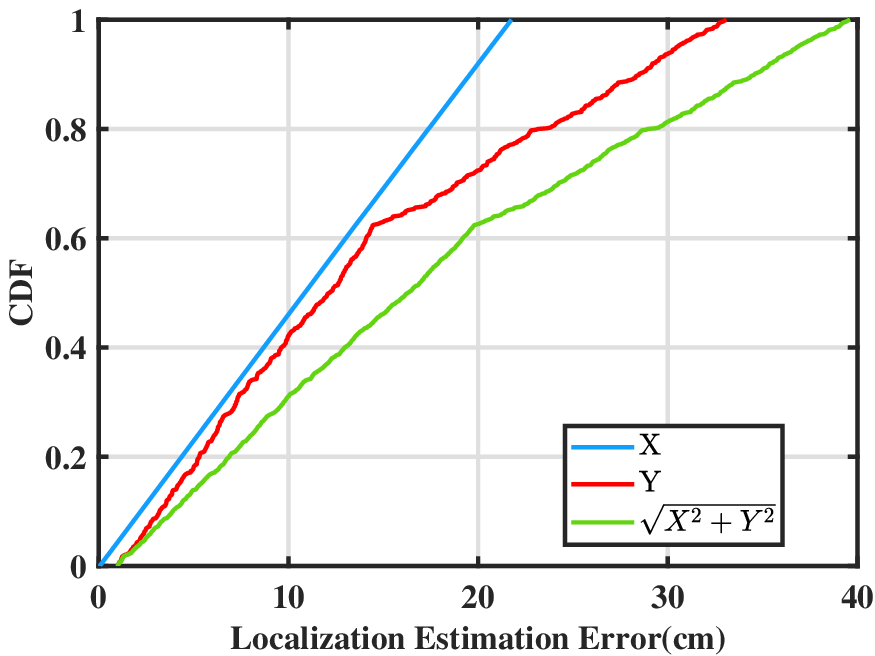}
\caption{Localization error for\\ lab.}
% \vspace{-0.4 cm}
\label{cdf_distance_lab}
\end{minipage}
\begin{minipage}{5.7cm}
\centering
\includegraphics[width=1\textwidth]{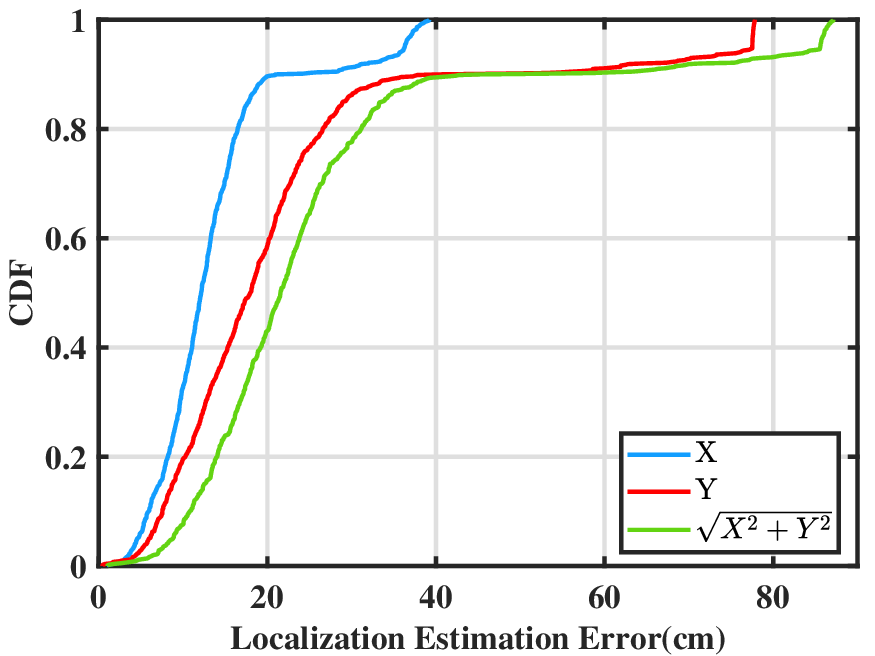}
\caption{Localization error for\\ around-the-corner.}
% \vspace{-0.4 cm}
\label{cdf_distance_conner}
\end{minipage}
% \vspace{-0.5 cm}
\end{figure*}

{
\sn was deployed in three different indoor environments for performance evaluation: (1) a laboratory, (2) an office, and (3) an around-the-corner scenario, as illustrated in Fig.~\ref{experimental_setup}. In each configuration, the radar is placed on one side of a reflector (wooden wall or door), while the target, equipped with a single tag, is positioned on the opposite side, outside the radar’s field of view, creating a non-penetrable NLoS scenario.  
Both the tag (attached to the target) and the radar are mounted at a height of 1.4 meters. The radar-reflector-target distance varies between 2m and 7m, with target positions indicated by red dots. Additionally, target orientations range from $-30^{\circ}$ to $30^{\circ}$ in $5^{\circ}$ increments for performance evaluation. Localization accuracy is assessed using position errors in Cartesian coordinate system along both X and Y directions as performance metrics.

}
%We use both distance (cm) and angle (degree) errors as metrics to evaluate 

\vspace{-0.3cm}
\subsection{Overall Performance}

\label{Overall Performance}

Fig. \ref{cdf_distance_office}, \ref{cdf_distance_lab}, and \ref{cdf_distance_conner} show
\sn' overall localization performance, i.e., the cumulative distribution functions (CDF) of estimation errors in X and Y directions, as well as the Euclidean distance error (i.e., $\sqrt{X^2+Y^2}$), in Office, Lab, and Around-the-corner environments, respectively.
These figures reveal two key findings. First, the proposed \sn achieves reliable localization performance despite using only a single tag, limited bandwidth, and a small number of antennas. The median errors for X, Y, and $\sqrt{X^2+Y^2}$ are $10.69,\mathrm{cm}$, $11.98,\mathrm{cm}$, and $15.89,\mathrm{cm}$ in the Lab environment; $14.25,\mathrm{cm}$, $21.62,\mathrm{cm}$, and $24.33,\mathrm{cm}$ in the Office environment; and $12.18,\mathrm{cm}$, $17.96,\mathrm{cm}$, and $21.51,\mathrm{cm}$ in the Around-the-Corner environment, respectively.  This is credited to our developed HFD design and FS-MUSIC, which enhance SNR and improve location resolution and thus contribute to accurate localization results. Therefore, the proposed \sn shows significant potential for real-world applications, especially in cases with limited available resources in terms of tag, bandwidth, and/or antenna. 

Second, \sn achieves the most favorable results in Lab among the three scenarios. The reason is that the wireless environment in Lab is relatively simpler (e.g., less cluttering, see Fig. \ref{experimental_setup}) than the other two configurations, which is beneficial for extracting valuable localization information. Unsurprisingly, \sn produces the worst performance in the most cluttering environment, i.e., Office. Although complex configurations may introduce multiple paths, which potentially benefits localization for \sn, the overall SNR condition is still a constraint on performance.

\subsection{Key Algorithm Performance} 
\label{Key Algorithm Performance} 
%In this section, we investigate the impact of \sn' key algorithms on localization performance. In this (i.e.,~\ref{Key Algorithm Performance}) and the following (i.e., \ref{Robustness Analysis}) sections, we will focus on the performance evaluation in the Office environment, and omit the results for Lab and Around-the-corner for brevity. We note that similar trends were 
%observed in Lab and Around-the-corner environments.

In this section, we analyze the impact of \sn's key algorithms on localization performance. For brevity, this section (\ref{Key Algorithm Performance}) and the next section (\ref{Robustness Analysis}) focus on evaluations conducted in the office environment, omitting results from the laboratory and around-the-corner scenarios. However, similar performance trends were observed in those environments.

\begin{figure*}[t]
\centering
\begin{minipage}{5.75cm}
\centering
\includegraphics[width=1\textwidth]{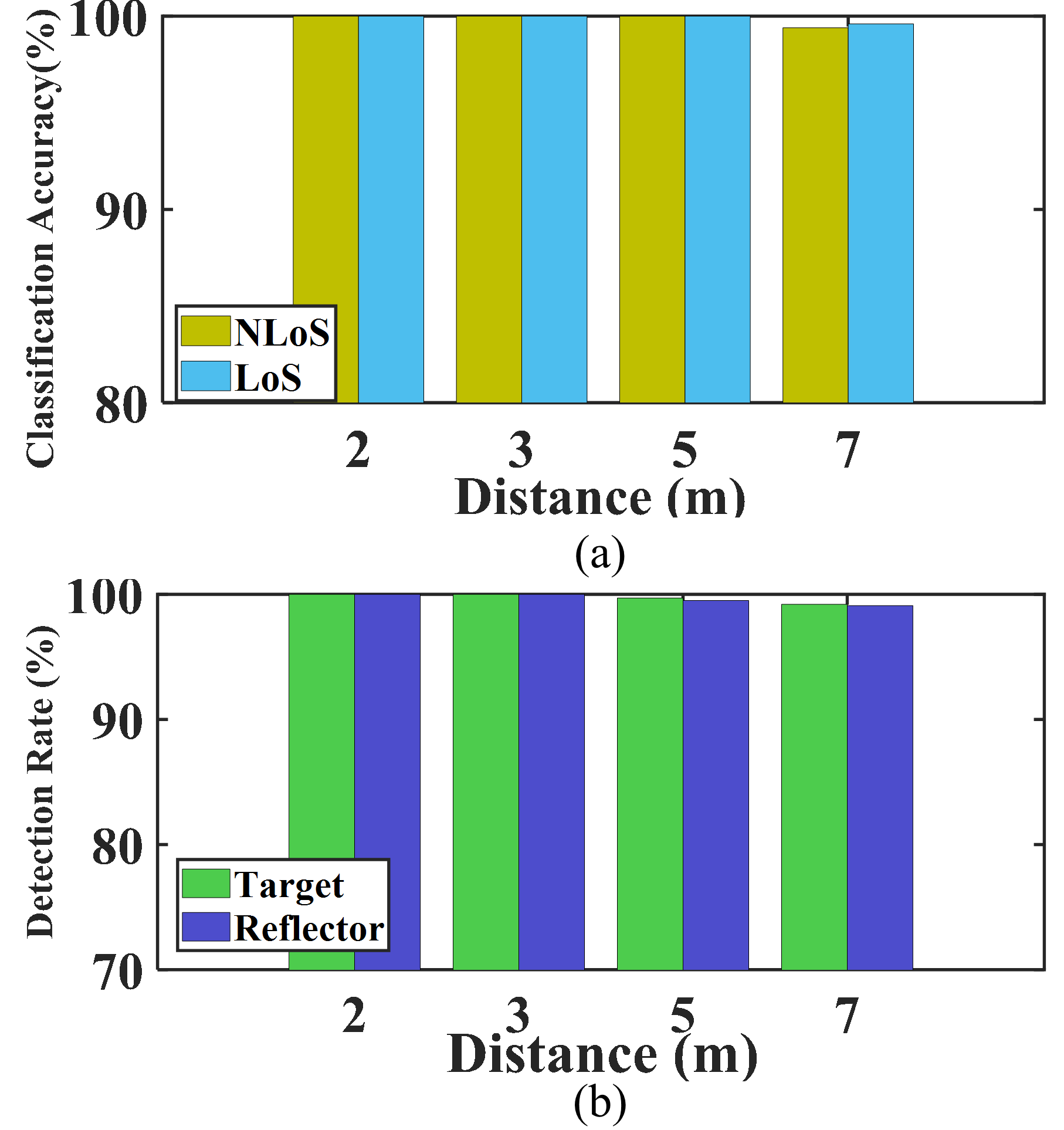}
\caption{Performance of classification and detection.}\label{robust_detection_rate}
\end{minipage}
\begin{minipage}{5.75cm}
\centering
\includegraphics[width=1\textwidth]{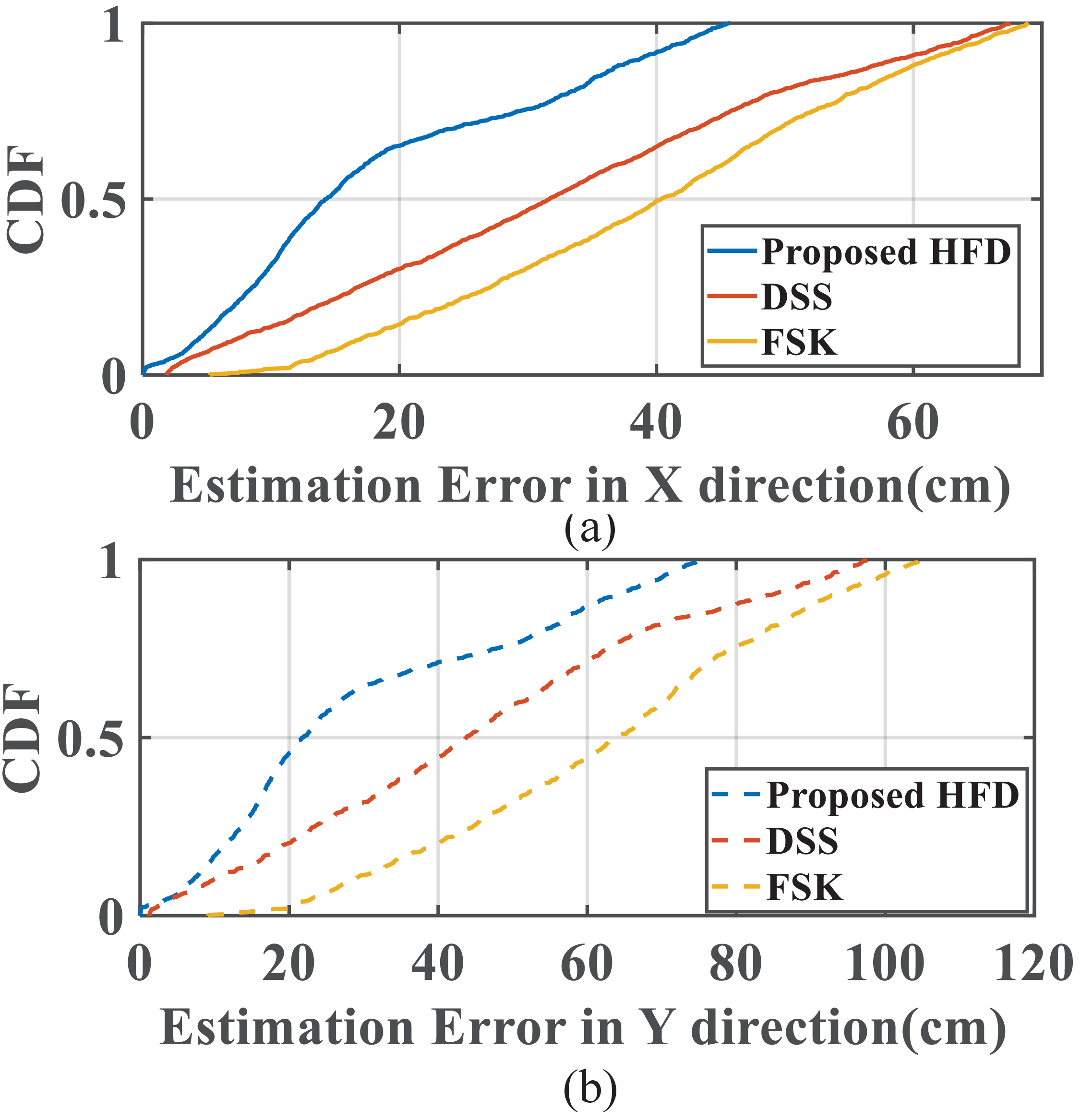}
\caption{Comparison of different modulation methods.}
% \vspace{-0.4 cm}
\label{algorithm_cdf_distance_office_FH}
\end{minipage}
\begin{minipage}{5.75cm}
\centering
\includegraphics[width=1\textwidth,height=6cm]{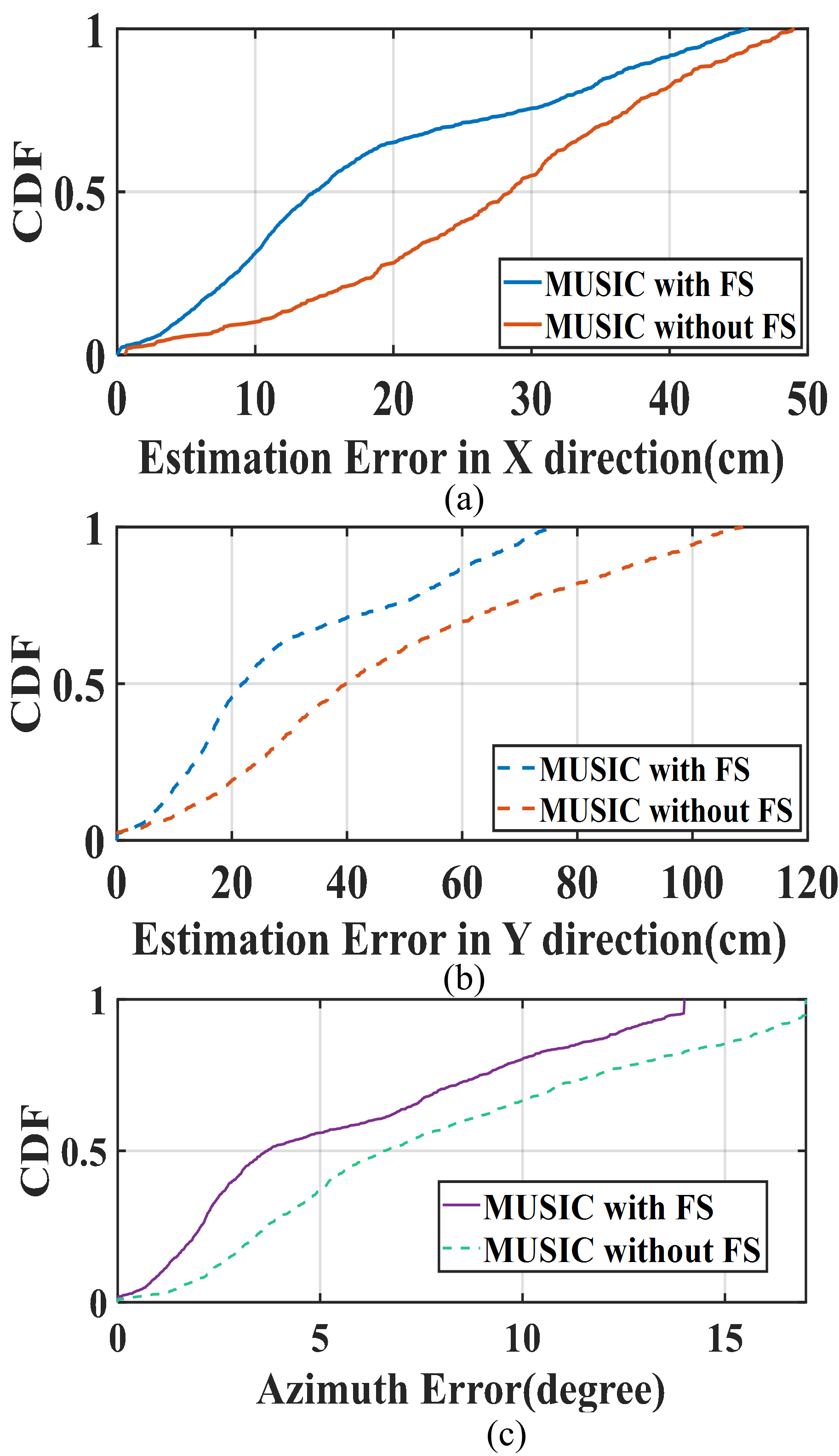}
\caption{Effectiveness of FS-MUSIC.}
% \vspace{-0.4 cm}
\label{algorithm_cdf_distance_office_MUSIC}
\end{minipage}
% \vspace{-0.3 cm}
\end{figure*}

\subsubsection{Effectiveness of HFD Modulation} \label{relector_detection}
{

Fig.~\ref{robust_detection_rate} presents the performance of the proposed HFD modulation in distinguishing between LoS and NLoS conditions, as well as in detecting the target and reflectors. The results indicate that classification and detection accuracy exceed 99\% for both the target and reflectors, even at distances of up to 7 meters. This high accuracy ensures reliable positioning and robust performance in various indoor environments, making HFD a practical choice for real-world applications.

Fig.~\ref{algorithm_cdf_distance_office_FH} compares localization results for different modulation techniques based on median errors in the X and Y directions. The proposed HFD design significantly outperforms conventional modulation schemes, including FSK and DSSS, in localization accuracy. Specifically, HFD achieves a median error of 14.25 cm in the X direction, compared to 40.32 cm for FSK and 31.6 cm for DSSS. Similarly, in the Y direction, HFD yields a median error of 21.62 cm, substantially lower than FSK (63.88 cm) and DSSS (40.79 cm). This improvement is attributed to HFD's combined use of FHSS and DSSS, which enhances interference mitigation and improves SNR conditions. Additionally, FHSS in the switching code design provides an extra domain for data sources, increasing measurement dimensions and contributing to accurate localization results (see Section~\ref{Effectiveness of MUSIC conjugate} below for further details).

}
\subsubsection{Effectiveness of FS-MUSIC} 
\label{Effectiveness of MUSIC conjugate}

{

Fig.~\ref{algorithm_cdf_distance_office_MUSIC} illustrates the impact of FS-MUSIC on estimation errors. The results demonstrate that FS-MUSIC significantly outperforms the conventional MUSIC method in both position and angle estimation. The median errors for FS-MUSIC in the X and Y directions are 14.25 cm and 21.62 cm, respectively—reductions of 50\% and 54\% compared to traditional MUSIC (28.46 cm and 39.98 cm). Furthermore, FS-MUSIC achieves a median angle estimation error of 3.79$^\circ$, which is 58\% lower than the 6.63$^\circ$ error of traditional MUSIC. This improvement is due to FS-MUSIC leveraging both spatial and frequency domains, whereas conventional MUSIC relies solely on the spatial domain. By increasing the number of measurement dimensions, FS-MUSIC extracts more valuable localization information, resulting in superior accuracy.

}

\subsection{Robustness Analysis}
\label{Robustness Analysis}

\begin{figure}[t]
\centering
\includegraphics[width=8.5 cm]{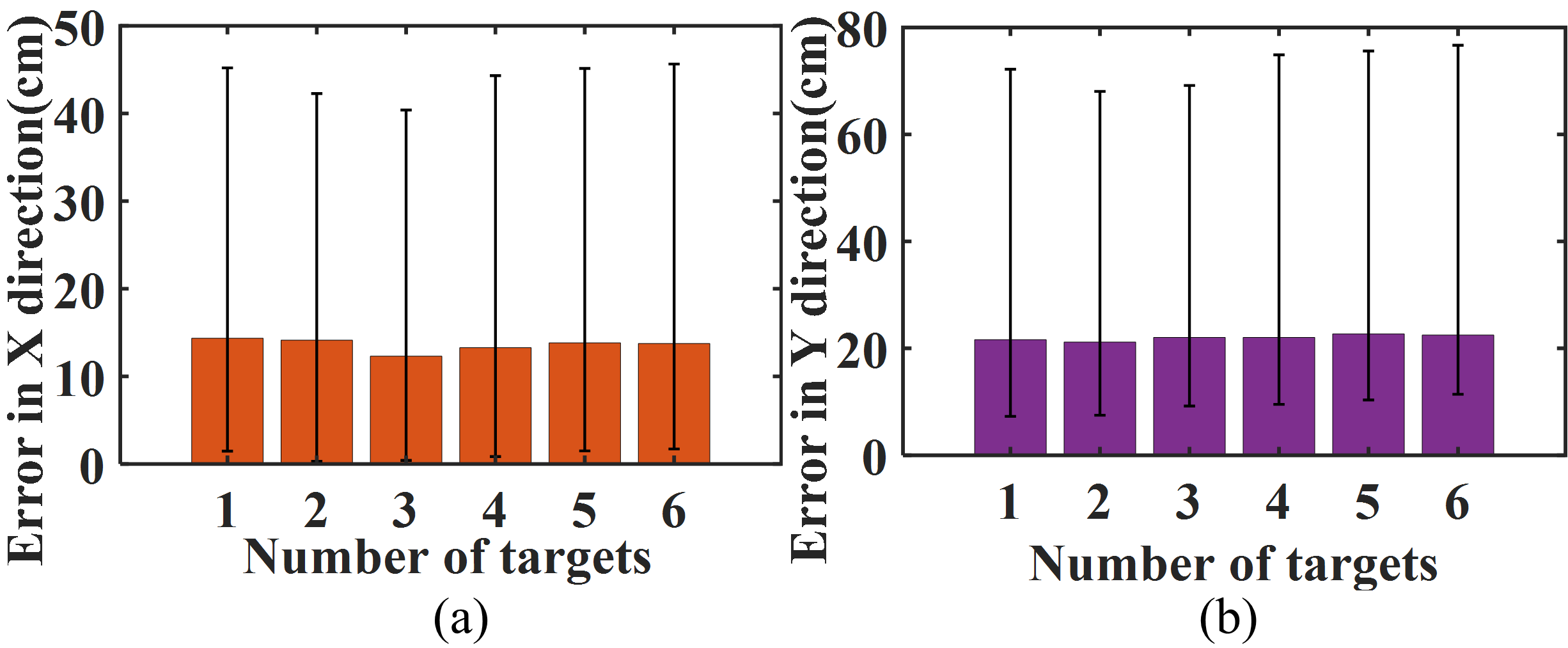 }
\DeclareGraphicsExtensions.
\vspace{-0.6 cm}
\caption{Impact of the number of targets.}
\vspace{-0.5 cm}
\label{robust_imp_tag}
\end{figure}

{
\subsubsection{Impact of Multi-Target Localization}
Fig.~\ref{robust_imp_tag} evaluates \sn's performance in a multi-target scenario, where each target is equipped with a unique tag. The results show that increasing the number of targets does not degrade localization accuracy in either the X or Y direction. This demonstrates that \sn can effectively localize multiple targets simultaneously. This robustness is achieved through the unique TLC sequence assigned to each tag, which leverages FHSS and DSSS to mitigate signal interference between targets. By ensuring minimal cross-correlation between sequences, \sn maintains high localization accuracy, making it well-suited for real-world multi-target deployments.

}

\subsubsection{Impact of different distances}
\begin{figure}[t]
\centering
\includegraphics[width=8.8 cm]{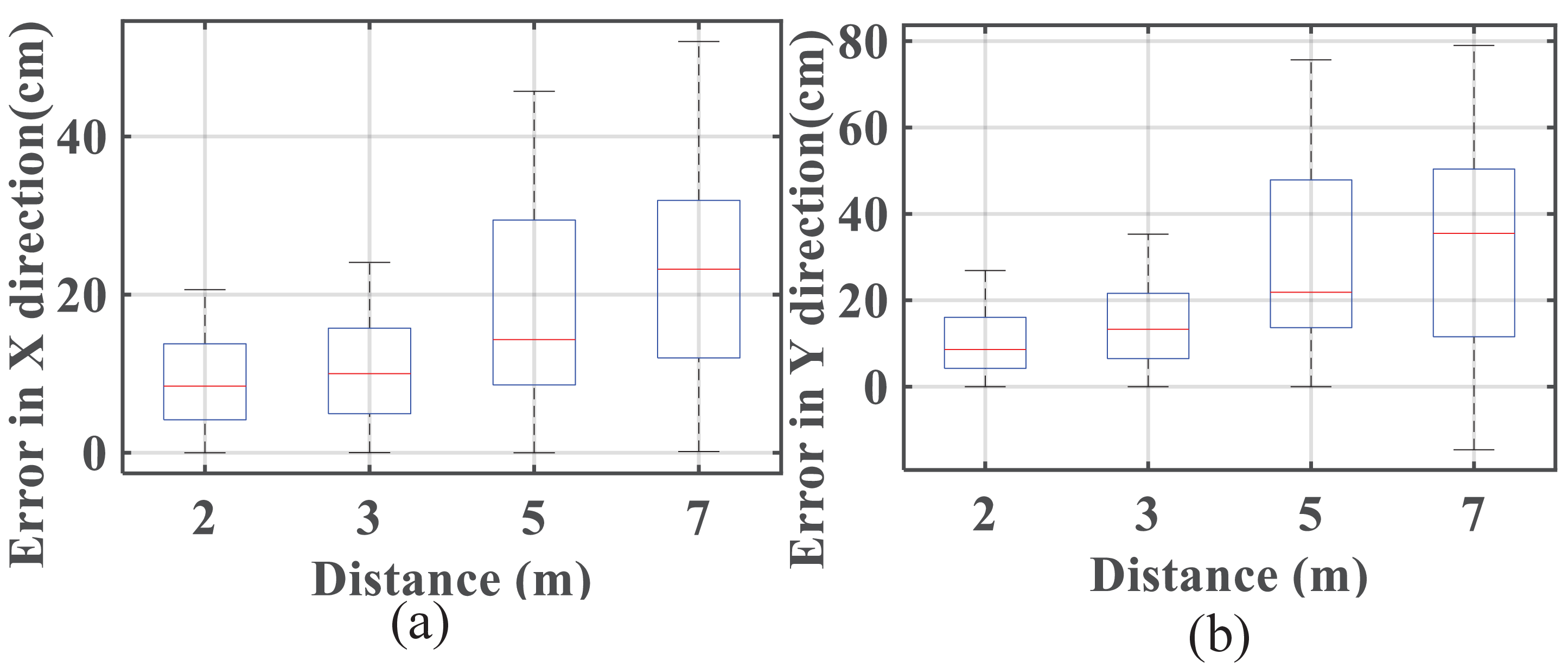 }
\DeclareGraphicsExtensions.
\vspace{-0.6 cm}
\caption{Impact of different distance.}
\label{robust_imp_distance_range}
\vspace{-0.3 cm}
\end{figure}

Fig.~\ref{robust_imp_distance_range} examines the effect of varying distances on \sn's performance. The radar-reflector-target distances tested are 2, 3, 5, and 7 meters. The results indicate that localization errors in both the X and Y directions increase as distance increases due to lower SNR conditions. Nevertheless, even at a measurement distance of 7 meters, \sn achieves errors below 30 cm in the X direction and 40 cm in the Y direction, demonstrating its potential for real-world indoor applications.

{
\subsubsection{Impact of Different Reflector Materials} \label{materials}
Table~\ref{IDR} evaluates the impact of different reflector materials on \sn's localization performance. Four common indoor materials are considered: metal, wood, concrete, and plaster. The results show that \sn can operate effectively across a wide range of reflectors, \textbf{including non-metallic surfaces}.

Reflectors with lower signal absorption, such as metal and concrete, result in better localization accuracy, with mean errors of 7.4 cm (X) and 11.2 cm (Y) for metal, compared to 16.6 cm (X) and 19.4 cm (Y) for plasterboard. These variations highlight \sn's adaptability to different environmental conditions. The system's robustness across diverse materials is primarily attributed to HFD modulation, which enhances localization accuracy even in suboptimal reflector conditions.

}

\begin{table}[tb]
\centering
\small
\caption{Impact of different reflector materials}
\vspace{-0.4 cm}
\begin{tabular}{|p{1.8cm}|>{\centering\arraybackslash}p{2.8cm}|>{\centering\arraybackslash}p{2.8cm}|}
\hline
\textbf{Materials} & \textbf{Mean Error in X Direction (cm)} & \textbf{Mean Error in Y Direction (cm)} \\
\hline
\text{Metal} & 7.4 & 11.2 \\
\hline
\text{Wood} & 11.2 & 13.7 \\       
\hline
\text{Concrete} & 9.9 & 16.7 \\
\hline
\text{Plaster} & 16.6 & 19.4 \\
\hline
\end{tabular}
\vspace{-0.4 cm}
\label{IDR}
\end{table}

\begin{table*} [tb]
\centering
\small
\caption{Comparison of existing NLoS localization methods}
\vspace{-0.2 cm}
\begin{tabular}{cccccc}
\hline
{ Method }  & SuperSight \cite{10.1145/3643832.3661857} & CornerRadar \cite{10.1145/3517226} &MetaSight \cite{10.1145/3498361.3538947} &\sn
\\
\hline
Localization precision   &Millimeter-level, $\le 8mm$    &Centimeter-level, $\le 20cm$  & Centimeter-level, $\le 15cm$  &Centimeter-level, $\le 11cm$ \\
\hline
Transmitter                &MMWCAS-RF-EVM &FMCW radar& ALN-9610&DEMORAD radar \\
Bandwidth     &3.89GHz & 1.8GHz  & 80 MHz &250MHz\\
Antenna    &$12\times 16$ & $12\times 12$  &$1\times1$&$2\times4$\\
\hline
Tag/Metasurface       &Tags (3) array & $-$  & Multiple meta surfaces & Single tag\\
Size      &$40cm\times40cm$ & $-$  & $105cm\times65cm$ & $12cm\times8cm$\\
\hline
Additional device       & $-$ & Lidar  & USRP and Impinj&  $-$\\
\hline
\end{tabular}
%\vspace{-0.2 cm}
\label{cmp_methods}
\end{table*}

\subsection{Power Consumption}
\label{Power Consumption}

The power consumption of \sn is driven by tag operations under the proposed HFD scheme, consisting of TLC and RLC phases as an effective duty cycle. In the TLC phase, the tag is active, and tag’s power usage is primarily due to modulation frequencies of 2 KHz, 5 KHz, and 10 KHz, with respective power consumptions of 0.348 mW, 0.373 mW, and 0.506 mW, resulting in an average of approximately 0.41 mW. In the RLC phase, the tag consumes no power due to inactive status. This results in an average power consumption of 41 $\mu$W across the full HFD sequence. With a CR2477 coin cell battery \cite{cell} with a 1,000 mAh capacity, the tag is estimated to operate continuously for approximately 5.6 years.

\section{Discussion and Future Work}
\label{Discussion and Future Work}

\textbf{Tag orientation and height}:  
The tag’s orientation and height significantly impact localization accuracy due to the directional nature of mmWave signals. Misalignment with the radar reduces signal strength and increases localization errors. In our setup, both the tag and radar are positioned at 1.4 meters to align with the antenna’s main lobe and minimize elevation-induced degradation. However, real-world deployments may introduce variations that affect antenna gain and signal reception. While our study evaluates tag orientations from $-30^{\circ}$ to $30^{\circ}$ in $5^{\circ}$ increments, future work could explore omnidirectional antennas and 3D antenna gain modeling to mitigate these effects.

\textbf{Reflector influence}:  Reflectors play a crucial role in successfully localizing the target in NLoS scenarios. However, when the target moves or the environment changes, the original reflectors may become blocked or ineffective, significantly degrading localization performance. The system developed in this work effectively addresses this concern through the unique design of HFD. Each modulated sequence in HFD is specifically crafted to accurately detect and distinguish reflectors from the target. As a result, \sn is particularly suitable for practical indoor applications, especially those involving mobile targets or changing environments.

%\vspace{-0.3cm}
\section{Related Work}
\label{sec:relatedWork}
% This section provides a comprehensive review of existing indoor localization techniques, emphasizing their properties, advantages, and limitations.

\noindent \textbf{Indoor Localization}: Extensive research has been devoted to indoor localization, resulting in a wide range of technologies. Optical localization is one of the most popular solutions, relying on visual and light-based systems such as cameras and LiDAR ~\cite{10.1145/3372224.3380894,10.1145/2461912.2461928,10.5555/3666122.3669255,8569569} to determine target positions. Although effective, optical localization faces challenges related to lighting conditions, privacy, and cost. RF-based localization has emerged as a promising alternative, utilizing variations in RF signals. WiFi and RFID are common choices for RF-based localization \cite{10.1145/3411834,10.1145/3130940,10.1145/3636534.3649355,10.1145/2348543.2348580,8884167,9546508,10.1145/3356250.3360019}, but they suffer from limitations like short operational range and low accuracy. In contrast, mmWave signals, with their high frequency and wide bandwidth, show promise for accurate and long-range localization \cite{10.1145/3241539.3241542,10.1145/3581791.3596869,10.1145/3386901.3389034,10.1145/3603165.3607428}. To achieve this, mmWave FWCW radars extensively employ FWCW signals for localization tasks.

\noindent \textbf{Van Atta Array}:
The VAA, consisting of an antenna configuration, plays a pivotal role in mmWave indoor localization \cite{1144877,trzebiatowski2022simple}. The core idea leverages retro-reflection, facilitated by the antenna reciprocity principle, to determine target location information \cite{bae2022omniscatter,8399826,10.1145/2348543.2348580,8309376}. This approach effectively mitigates interference and strengthens desired signals, enhancing localization accuracy and reliability. However, existing works are limited to linear VAA configurations, resulting in sharp radar cross-section (RCS) reductions during tag-radar elevation misalignment \cite{6359763,8115232}. To address this concern, planar VAA has gained interest and is widely used in mmWave localization systems to improve RCS stability, spatial diversity, and signal robustness \cite{5559365,6494356,lu2023millimeter,soltanaghaei2021millimetro}. Despite the progress, these methods primarily focus on LoS conditions, which do not fully capture the complexities of indoor environments characterized by NLoS scenarios. Consequently, their applicability in indoor environments is constrained, especially under non-penetrable NLoS conditions.

\noindent{\textbf{Non-penetrable NLoS localization}}: 
Non-penetrable NLoS localization holds increasing promise due to its unique advantages in maintaining robust and reliable indoor positioning. A recent solution proposed in \cite{10.1145/3517226} designs an indoor localization system by leveraging signals assisted by LiDAR devices. Similarly, the works in \cite{10.1145/3498361.3538947} and \cite{10.1145/3643832.3661857} propose localizing blocked objects using multiple metasurfaces and triangular tag arrays, respectively. Although these approaches can achieve localization in non-penetrable NLoS conditions, they rely on specialized equipment such as LiDAR, metasurfaces, or tag arrays, which escalates deployment costs and limits their application potential. In contrast, our proposed method (\sn) achieves accurate localization (at centimeter-level precision with 250 MHz bandwidth) using only a single tag per target. Table \ref{cmp_methods} summarizes the key differences between the proposed \sn against existing NLoS localization methods.

%\vspace{-0.2cm}
\section{Conclusion}
\label{Conclusion}
We introduce a novel mmWave system, \sn, tailored for accurate indoor non-penetrable NLoS localization using only a single radar and a tag alone. We propose an innovative modulation scheme for reflector and surroundings detection, distinguishing them from the target, through the hybrid utilization of FHSS and DSSS techniques. To enhance SNR levels in NLoS scenarios, we exploit the modulated signal's correlation property to mitigate the influence of noise and retain the desired signal. Moreover, we propose an innovative scheme to enhance localization accuracy by utilizing specifically designed frequency-spatial dimensions in conjunction with the MUSIC algorithm. Extensive evaluations are across various scenarios and conditions, and results demonstrate that \sn is capable of realizing accurate localization with solely a single tag and radar, offering promising potential for real-world applications.

\balance
\bibliographystyle{ACM-Reference-Format}
\bibliography{sample-base}

\appendix

\end{document}